\documentclass[aps,pra,reprint,showpacs,superscriptaddress,nofootinbib,twocolumn,longbibliography]{revtex4-2}
\usepackage{amsmath}
\usepackage{amsfonts}
\usepackage{amsthm}
\usepackage{amssymb}
\usepackage{graphicx}
\usepackage{enumerate}
\usepackage{color}
\usepackage{lineno}
\usepackage[T1]{fontenc}
\usepackage{mathptmx}
\usepackage{braket}
\usepackage{todonotes}
\usepackage{soul}
\usepackage{subfigure}
\graphicspath{{figures/}}

\DeclareMathOperator{\Tr}{Tr}

\newtheorem{theorem}{Theorem}

\newtheorem{corollary}{Corollary}

\newtheorem{observation}{Observation}

\usepackage{float}
\usepackage{mathrsfs,amsfonts,dsfont}
\usepackage{epsfig}
\usepackage{bm}
\usepackage{framed}
\usepackage{tcolorbox}
\usepackage{multirow}
\usepackage{verbatim}
\usepackage{physics}
\usepackage[10pt]{moresize}
\usepackage{amscd}
\usepackage{amssymb}
\usepackage{tikz}
\usepackage[ruled,vlined]{algorithm2e}
\usepackage[keys,primitives]{cryptocode}
\usepackage{tabularx}
\usepackage{diagbox}
\usepackage{lipsum}
\usepackage[utf8]{inputenc}
\usepackage[bookmarks=false,colorlinks,citecolor=blue,
linkcolor=blue,anchorcolor=blue,urlcolor=blue
]{hyperref}

\begin{document}

\title{Certifying optimal device-independent quantum randomness in quantum networks}
\author{Shuai Zhao}
\email{zhaoshuai@hdu.edu.cn}
\affiliation{School of Cyberspace, Hangzhou Dianzi University, Hangzhou 310018,China}
\affiliation{Zhejiang Provincial Key Laboratory of Sensitive Data Security and Confidentiality Governance, Hangzhou 310018,China}
\author{Rong Wang}
\email{ericrongwang19@gmail.com}
\affiliation{School of Cyberspace, Hangzhou Dianzi University, Hangzhou 310018,China}
\affiliation{Zhejiang Provincial Key Laboratory of Sensitive Data Security and Confidentiality Governance, Hangzhou 310018,China}
\author{Qi Zhao}
\email{zhaoqcs@hku.hk}
\affiliation{QICI Quantum Information and Computation Initiative, School of Computing and Data Science,
The University of Hong Kong, Pokfulam Road, Hong Kong}





\begin{abstract}
  Bell nonlocality provides a device-independent (DI) way to certify quantum randomness, based on which true random numbers can be extracted from the observed correlations  without detail characterizations on devices for quantum state preparation and measurement. However, the efficiency of current strategies for DI randomness certification is still heavily constrained when it comes to non-maximal Bell values, especially for multiple parties. Here, we present a family of multipartite Bell inequalities that allows to certify optimal quantum randomness and self-test GHZ (Greenberger-Horne-Zeilinger) states, which are inspired from the stabilizer group of the GHZ state. Due to the simple representation of stabilizer group for GHZ states, this family of Bell inequalities is of simple structure and can be easily expanded to more parties. Compared with the Mermin-type inequalities, this family of Bell inequality is more efficient in certifying quantum randomness when non-maximal Bell values achieved. Meanwhile, the general analytical upper bound for the Holevo quantity is presented, and achieves better performance compared with the MABK (Mermin-Ardehali-Belinskii-Klyshko) inequality, Parity-CHSH (Clauser-Horne-Shimony-Holt) inequality and Holz inequality at $N=3$, which is of particular interests for experimental researches on DI quantum cryptography in quantum networks.
\end{abstract}
\maketitle
Intrinsic randomness is crucial for a lot of applications and has attracted extensive research interests ranging from cryptography to sociology \cite{colbeck2007quantum,senno2023quantifying,herrero2017quantum}. Due to the deterministic feature, it has been well established that one cannot certify intrinsic randomness within classical theories, and thus cryptographic protocols relying solely on classical theories in principle can be predicted~\cite{pironio2010random,Bell1964,goh2018geometry}. In contrast, intrinsic randomness is one of the fundamental features of quantum theories and forms the basis for many promising quantum information protocols~\cite{ladd2010quantum,gisin2002quantum,pezze2018quantum}, such as quantum key distribution~\cite{xu2020secure}, blind quantum computing~\cite{broadbent2009universal} and hybrid quantum-classical cryptography~\cite{li2023device}. Therefore, true random numbers can be extracted directly from quantum systems, such as that based on the process of splitting on a photon beam by a 50:50 splitter~\cite{jennewein2000fast}.

However, imperfections of device are inevitable in experiments and should be well distinguished from adversaries's attacking, which is very difficult to confirm due to the complexity of quantum systems~\cite{colbeck2007quantum,pironio2010random,pironio2013security}. To further release the assumptions on devices for preparing and measuring quantum states, Bell nonlocality is extensively adopted to certify intrinsic randomness in quantum systems, due to the fact that Bell nonlocal correlations cannot be described by any local realistic theories. Thus, once the Bell nonlocality correlation is observed in experiments, one can certify true randomness no matter the mechanism underlying the preparation and measurement devices, i.e. \emph{device-independent (DI)} quantum randomness certification\cite{colbeck2007quantum,pironio2010random}.

\begin{figure}
  \centering
  \vspace{0cm}
  \includegraphics[width=0.49\textwidth]{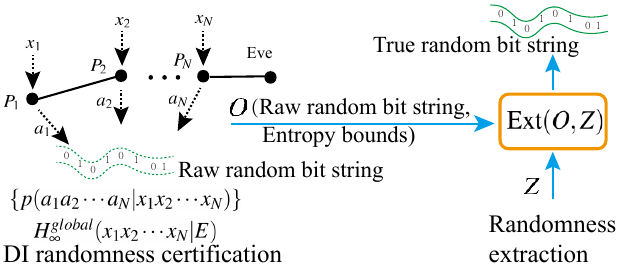}\\
  \caption{Diagram of a typical device-independent quantum random number generation protocol. The entropy bound is derived by the DI quantum randomness certification protocol (perhaps with the help of technics for multiple rounds experiments~\cite{arnon2018practical,zhang2018certifying,wang2025one}). Then, the true random bit string is extracted through the extractor $\text{Ext}(O,Z)$ with raw bit string and entropy bound as inputs $O$ by consuming a short random seed $Z$. $x_i$ and $a_i$ are input and output for the lab of party $P_i$. $\{p(a_1a_2\cdots a_N|x_1x_2\cdots x_N)\}$ is the probability distribution for the measurement results among parties $P_1$, $P_2$, $\cdots$, $P_N$. $H_{\infty}^{global}(x_1x_2\cdots x_N|E)$ denotes min-entropy for given measurement inputs $x_1, x_2,\cdots,x_N$ under quantum eavesdropping operation $E$. Black dots within $P_i$ and Eve are untrusted quantum systems. Blue arrows are directions for data flows indicating data inputs and outputs for the Extractor.}\label{Frame_Random_Generator}
\end{figure}

Specifically, a typical DI quantum random number expansion protocol can be separated into two parts: entropy source and extractor~\cite{herrero2017quantum,zhang2025randomness,amer2025applications,kavuri2025traceable}. As shown in Fig.~\ref{Frame_Random_Generator}, the raw random bit string is derived from the entropy source, which is composed of untrusted quantum systems. DI quantum randomness certification pursues certifying quantum randomness more efficiently in untrusted quantum systems. Immediately, supervised by the certified entropy bound, the raw random bit string is processed into fine ture random bit string in DI quantum random number expansions. So far, it's been demonstrated that true randomness can be certified in quantum systems even under no-signalling eavesdropping strategies~\cite{pironio2010random,zhao2023tilted}. Therefore, device-independent quantum randomness certification plays an essential role in DI quantum random number expansions with untrusted systems. Besides DI quantum random number expansions, DI quantum randomness certification has also laid the foundations for DI quantum randomness amplification, which takes weak randomness in measurement setting choices into consideration~\cite{colbeck2012free,gallego2013full,ramanathan2016randomness,xu2018improved,kessler2020device,zhao2023tilted,foreman2023practical}.

Typically, for a $(N,m,d)$ Bell scenario involving $N$ party quantum systems (see Fig.~\ref{Frame_Random_Generator}), each party $P_i$ ($i\in\{1,2,\cdots,N\}$) makes measurement choices with $x_i\in\{\hat{A}_{i0}, \hat{A}_{i1},\cdots,\hat{A}_{im}\}$ and obtains outcomes with $a_i\in\{0,1,\cdots, d-1\}$~\cite{dhara2013maximal}. The measurement results can be summarized as a probability distribution $p(a_1a_2\cdots a_N|x_1x_2\cdots x_N)$, in which the optimal global randomness implied is $N \log_2 d$ bits for projective measurements~\cite{dhara2013maximal,farkas2026maximal}. Without loss of generality, the global quantum randomness in Bell scenarios can be quantified by the min-entropy
\begin{equation}
  H_{\infty}^{\text{global}}(x_1x_2\cdots x_N)=-\log_2 G(x_1x_2\cdots x_N),
\end{equation}
where $G(x_1x_2\cdots x_N)=\max_{a_1,a_2,\cdots,a_N}p(a_1a_2\cdots a_N|x_1x_2\cdots x_N)$ is the guessing probability for given measurement settings $x_1,x_2,\cdots, x_N$. Similarly, the local quantum randomness is quantified by the guessing probability of marginal outcomes of each parties. Then, one can define the maximal global randomness over any measurement settings in Bell scenarios as  $H_{\infty}^{\text{global}}=\max_{x_1,x_2,\cdots, x_N}H_{\infty}^{\text{global}}(x_1x_2\cdots x_N)$. So far, the research on DI randomness certification with projective measurements on binary quantum systems has been pushed into regimes of (1) investigating the trade-off relation between Bell nonlocality and quantum randomness~\cite{acin2012randomness,wooltorton2023expanding}, (2) identifying more efficient Bell inequalities which offer the highest possible randomness at maximal Bell values~\cite{dhara2013maximal,de2015maximally,wooltorton2023expanding}, and (3) robustness analysis on the certified randomness versus non-maximal violation of Bell inequalities~\cite{bancal2014more,nieto2014using,nieto2018device,woodhead2018randomness,grasselli2023boosting}. (4) to certify quantum randomness under weak random sources~\cite{kessler2020device,zhao2023tilted}, (5) analysis on von Neumann entropy for DI cryptographic applications~\cite{ribeiro2019reply,holz2020genuine,grasselli2023boosting}, etc.

Along these directions, it has been identified that optimal $N$ bits of global randomnes can be certified in $(N,2,2)$ Bell scenarios for $N\geq 2$~\cite{acin2012randomness,dhara2013maximal,de2015maximally,acin2016optimal,woodhead2018randomness,woodhead2020maximal,wooltorton2022tight,zhao2023tilted,wooltorton2023expanding}. Moreover, it's been confirmed that the certified quantum randomness is an inequivalent resource to Bell nonlocality in two party and multiparty cases~\cite{acin2012randomness,wooltorton2022tight,wooltorton2023expanding}. Besides, the randomness certification has also been pushed to full range of weak random sources, which is important for DI randomness amplification protocols\cite{kessler2020device,zhao2023tilted,kulikov2024device}. However, on the research of (2) and (3), as illustrated in Table~\ref{table:comparison}, the robustness analysis is only conducted up to $N=3$ case for Mermin type Bell inequality. Meanwhile, on the research of (5), as in Fig.~\ref{comparision:grasselli}, the one-outcome von Neumann entropy based on the MABK inequality~\cite{mermin1990extreme,ardehali1992bell,belinskiui1993interference}, Parity-CHSH inequality~\cite{ribeiro2019reply} and Holtz inequality~\cite{holz2020genuine,grasselli2023boosting} are brilliant and urgent for improvements. Therefore, to construct efficient multipartite Bell inequalities and investigate its robustness performance under non-maximal Bell values is of great interests for both theoretical and experimental researches.

In this work, to further facilitate investigations on DI quantum randomness certifications, we construct a new family of Bell inequalities according to the stabilizer group of the GHZ state~\cite{greenberger1989going}(see \textbf{Theorem}~\ref{prop:1}). It can certify optimal quantum randomness for arbitrary $N$ parties at its maximal Bell values by taking $\alpha\rightarrow \infty$. To investigate the robustness performance for randomness certification under non-maximal Bell values, we conduct the robustness analysis for $N=3, 4$ parties and make a comparison with the Mermin inequality using the NPA hierarchy method~\cite{navascues2007bounding,navascues2008convergent}. The results show that compared with Mermin inequality in Ref.~\cite{woodhead2018randomness}, the family of multipartite Bell inequality in this work is more efficient for DI randomness certification at most range of non-maximal Bell violation (see Fig.~\ref{compare}) when $N=3$. Meanwhile, we obtained the analytical upper bound of the Holevo quantity, which also performances better compared with the MABK inequality, Parity-CHSH inequality and Holz inequality (see Fig.~\ref{comparision:grasselli} ). Furthermore, this family of Bell expression keeps the merit that the complexity increases polynomially with $N$ scaling up, which can be expanded to more parties easily. Due to the fact that the correlators involved in this family of Bell inequality are two-party correlation besides the first term, it's also of simple structure. Need to note that, this stabilizer-type method for constructing multipartite can also be generalized to other multipartite entanglement states with stabilizer representations besides the GHZ state.

\begin{table}[h]
  \caption{Comparison with works that certifying optimal $N$ bits quantum randomness. }\label{table:comparison}
  \begin{tabular*}{0.45\textwidth}{@{\extracolsep{\fill}} c c c c c c}
  \hline\hline
  \multirow{2}*{Works}  &\multirow{2}*{Year}&\multicolumn{2}{c}{Randomness(bits)}&\multirow{2}*{Robustness}\\
  \cline{3-4}& &Local & Global & analysis\\
  \hline
  Ref.~\cite{dhara2013maximal}&2013 & 1  & $N$(for odd $N$) & Null\\
  Ref.~\cite{de2015maximally}&2015 & 1 & $N$  & Null\\
  Ref.~\cite{woodhead2018randomness}&2018 & 1 & 3  & $N=3$ \\
  Ref.~\cite{wooltorton2023expanding}& 2023 & 1 & $N$ & Null \\
  This work  &2026 & 1 & $N$ (for $\alpha\rightarrow \infty$)  & $N=3,4$ \\
  \hline\hline
 \end{tabular*}
\end{table}


\section{\textbf{Constructing multipartite Bell inequalities}} To construct multipartite Bell inequalities for certifying optimal DI quantum randomness, one effective way is to expanding from the two party Bell inequality. Along this routine, the generalized Mermin-type expansion is extensively adopted and can certify optimal global randomness~\cite{dhara2013maximal,de2015maximally}(see Table~\ref{table:comparison}). Another type expansion is partition-type construction, which can certify optimal global randomness, too~\cite{curchod2019versatile,wooltorton2023expanding}.

Here, we present a family of multipartite Bell inequalities expanded from the $\alpha$-CHSH expression $\hat{I}_{\alpha}=\alpha \hat{A}_{10}\hat{A}_{20}+\hat{A}_{10}\hat{A}_{21}+\hat{A}_{11}\hat{A}_{20}-\hat{A}_{11}\hat{A}_{21}$
which was introduced in Refs.~\cite{bancal2014more,zhao2023tilted,wooltorton2023expanding}. When $\alpha=\frac{1}{2}$, the $\alpha$-CHSH expression $\hat{I}_{\alpha=\frac{1}{2}}$ can certify 2 bits of quantum randomness at its maximal quantum value which is optimal for $(2,2,2)$ Bell scenarios. Inspired by Refs.~\cite{baccari2020scalable,zhao2022constructing}, this new family of multipartite Bell inequalities is stabilizer-type construction for the GHZ state, in which the number of correlators increase polynomially with $N$ (see Appendix \ref{counstruction_stabilizer} for details). Specifically, the $N$-party Bell expression is of the following form:
\begin{equation}\label{Bell_expression}
  \hat{B}_{N}=(\hat{A}_{10}+\hat{A}_{11})\hat{A}_{20}\cdots \hat{A}_{N0}+\sum_{i= 2}^{N}(\alpha \hat{A}_{10}-\hat{A}_{11})\hat{A}_{i1},
\end{equation}
where \{$\hat{A}_{i0}$, $\hat{A}_{i1}$\} is the measurement setting with binary outcomes for the $i$-th party, $\alpha$ is an adjustable parameter. For $N\geq 3$, the upper bounds of the classical theory is
\begin{equation}
    B_{N}^{LHV}=\left\{
    \begin{array}{ll}
      2-(N-1)(\alpha-1), & \hbox{$\alpha\in(\alpha_L, \frac{1}{N-1}]$;} \\
      (N-1)(\alpha+1), & \hbox{$\alpha\in(\frac{1}{N-1},\infty)$.}
    \end{array}
  \right.
\end{equation}
where $\alpha_L=\frac{2N^2-2N\sqrt{N^2-2N+2}+N-1}{4N^2-5N+1}$ is the critic point indicating quantum Bell nonlocality correlation. While, for quantum theory, we have following results:
\begin{theorem}\label{prop:1} \textbf{Upper bound and self-testing statements}.

  The upper bound for the Bell expression in Eq.~\ref{Bell_expression} is
\begin{equation}
    B_{N}^{Q}=\sqrt{1+(N-1)^2\alpha^2}+\sqrt{1+(N-1)^2},
\end{equation}
which can be realized and up to local isometries, by the $N$-party GHZ state
\begin{equation}\label{ghz_state}
  |GHZ\rangle=\frac{1}{\sqrt{2}}(|00\cdots0\rangle+|11\cdots1\rangle),
\end{equation}
and the measurements
\begin{equation}\label{observable1}
  \begin{split}
    \hat{A}_{10}&=\frac{(N-1)\alpha}{\sqrt{1+(N-1)^2\alpha^2}}\sigma_z+\frac{1}{\sqrt{1+(N-1)^2\alpha^2}}\sigma_x,\\
    \hat{A}_{11}&=-\frac{N-1}{\sqrt{1+(N-1)^2}}\sigma_z+\frac{1}{\sqrt{1+(N-1)^2}}\sigma_x.
  \end{split}
\end{equation}
and
\begin{equation}\label{observable2}
  \begin{split}
    \hat{A}_{i0}=\sigma_x,\qquad \hat{A}_{i1}&=\sigma_z.
  \end{split}
\end{equation}
\end{theorem}
The proof is presented in Appendix \ref{proof_theorem}.



\begin{figure}[ht]
  \centering
  \includegraphics[width=.98\linewidth]{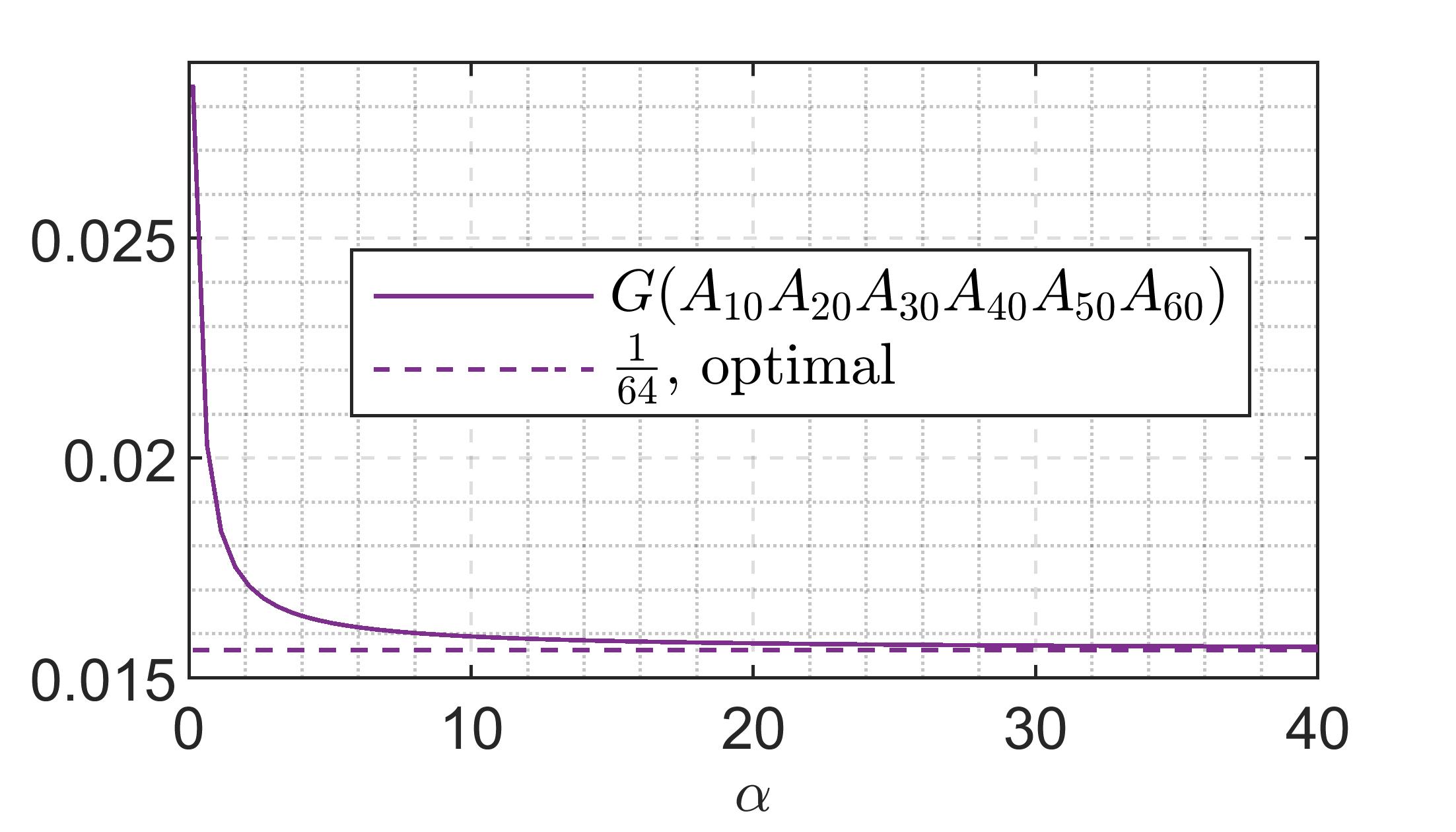}
  \caption{Guessing probability versus $\alpha$ at $N=6$ when maximal Bell values achieved. By adjusting $\alpha$, the guessing probability approaches the optimal values which are indicated by dashed line.}
  \label{figGuess}
\end{figure}

\section{\textbf{Certifying Optimal Global Quantum Randomness}} For $N=2$ case, it has been established that when $\alpha=\frac{1}{2}$ the $\alpha$-CHSH expression can certify the optimal 2 bits of quantum randomness\cite{zhao2023tilted,wooltorton2023expanding}. Here, we present that for $N\geq 3$ case, the above multipartite Bell expression $\hat{B}_{N}$ can certify optimal quantum randomness. The details are presented as follows:
\begin{corollary}
  It can be demonstrated from Eqs.~\ref{observable1} and \ref{observable2} that when $\alpha\rightarrow \infty$, the optimized observables are
\begin{equation}
  \begin{split}
    \hat{A}_{10}&=\frac{(N-1)\alpha}{\sqrt{1+(N-1)^2\alpha^2}}\sigma_z+\frac{1}{\sqrt{1+(N-1)^2\alpha^2}}\sigma_x \\
    &\rightarrow \sigma_z,\\
    \hat{A}_{11}&=-\frac{N-1}{\sqrt{1+(N-1)^2}}\sigma_z+\frac{1}{\sqrt{1+(N-1)^2}}\sigma_x, \\
     ~\hat{A}_{i0}&=\sigma_x,~\text{for}~i\geq 2, \qquad ~\hat{A}_{i1}=\sigma_z,~\text{for}~i\geq 2,
  \end{split}
\end{equation}
which can certify optimal $N$ bits of global randomness
\begin{equation}
  H_{\infty}^{\text{global}}=-\log_2 G(\hat{A}_{10}\hat{A}_{20}\cdots \hat{A}_{N0}) \xrightarrow{\alpha\rightarrow \infty} N,
\end{equation}
with
\begin{equation}
  G(\hat{A}_{10}\hat{A}_{20}\cdots \hat{A}_{N0})=\frac{1}{2^N}(1+\frac{1}{\sqrt{1+(N-1)^2\alpha^2}}),
\end{equation}
using the quantum states in Eq.~\ref{ghz_state}.

Meanwhile, it can also be derived that 1 bits of local randomness can be certified
\begin{equation}
  H_{\infty}^{\text{local}}=H_{\infty}^{\text{local}}(\hat{A}_{i1})=1, ~\forall i \geq 2.
\end{equation}
\end{corollary}

For example, when $N=3$, the guessing probability in measurement settings $\hat{A}_{10}\hat{A}_{20}\hat{A}_{30}$ is
\begin{equation}
  G(\hat{A}_{10}\hat{A}_{20}\hat{A}_{30})=\frac{1}{8}(1+\frac{1}{\sqrt{1+4\alpha^2}})\xrightarrow{\alpha\rightarrow \infty} \frac{1}{8}.
\end{equation}
It indicates the corresponding randomness approaches $3$ bits which is optimal for three party quantum systems.  To be more general, the guessing probabilities versus $\alpha$ at $N=6$ are plotted in Fig.~\ref{figGuess}, which are calculated at maximal Bell values. Need to note that the optimal DI quantum randomness is certified by taking $\alpha\rightarrow \infty$, which is interesting from the theoretic perspective. While, in experimental researches, it's practical to take finite $\alpha$ value considering loophole-free Bell tests.

\begin{figure}
  \centering
  \subfigtopskip=0pt
  \subfigbottomskip=0pt
  \subfigure[$N=3$]{\label{compare}
  \includegraphics[width=1\linewidth]{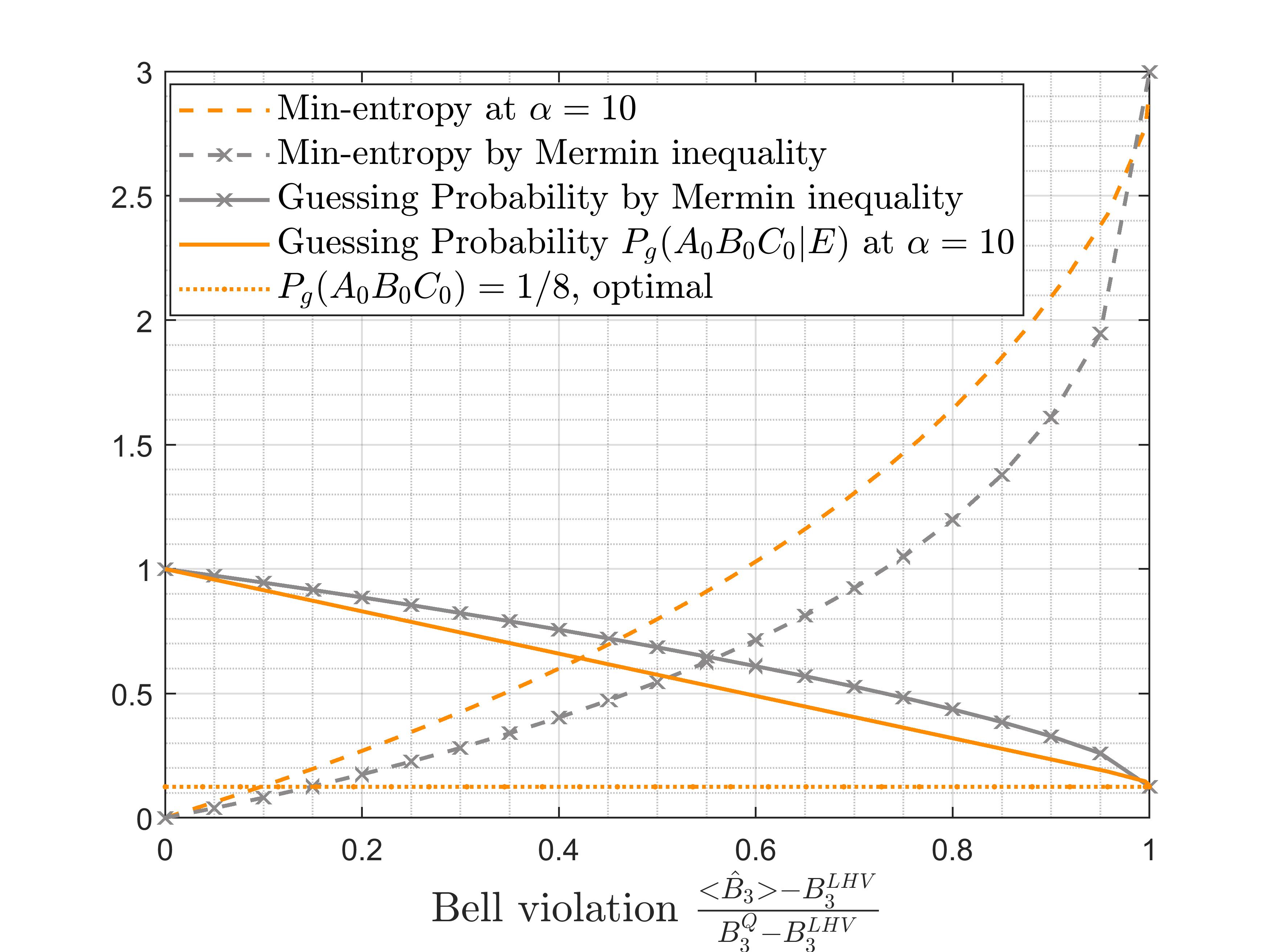}}
  \subfigure[$N=4$]{\label{performance_N=4}
  \includegraphics[width=1\linewidth]{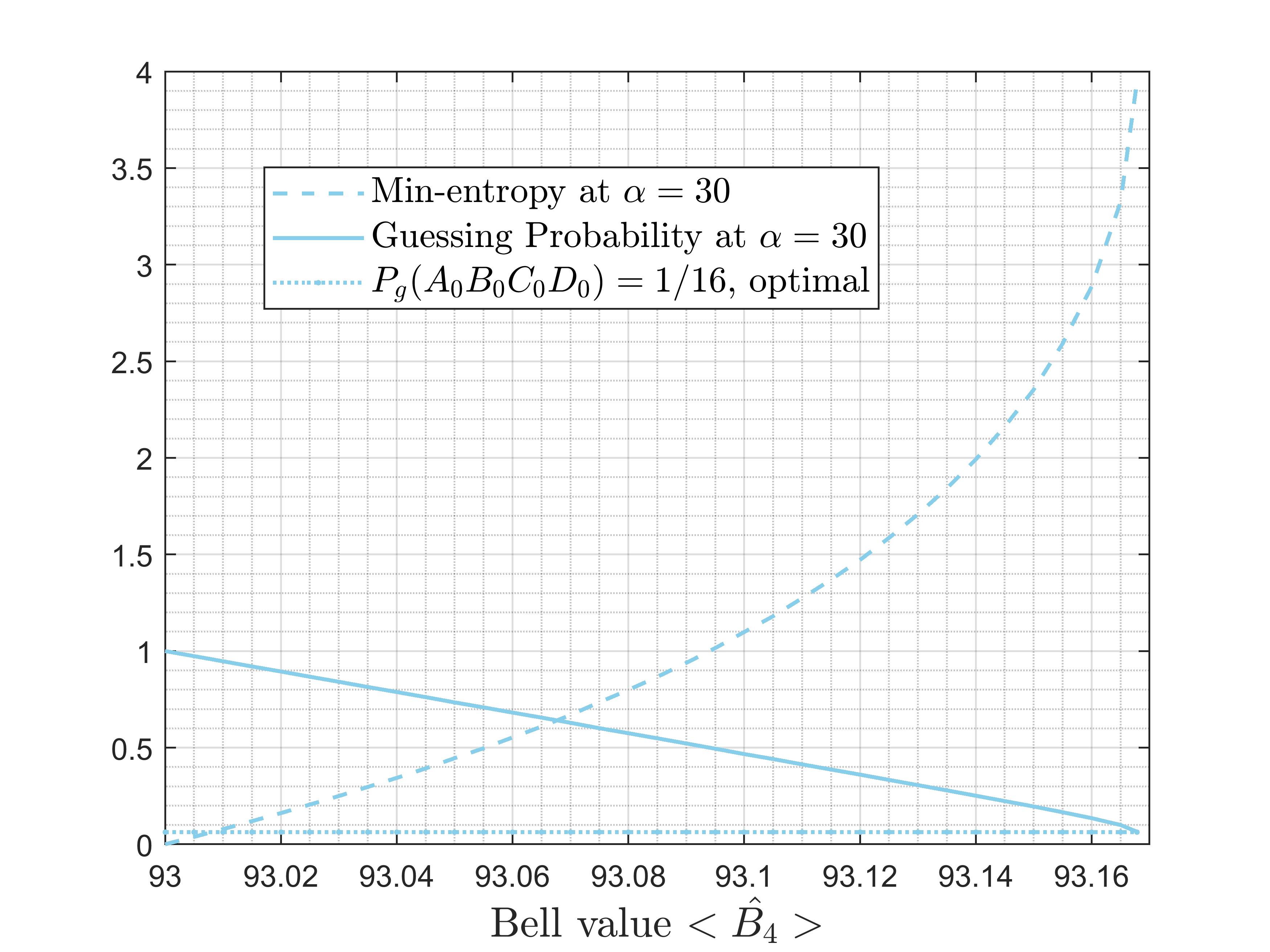}}
  \caption{Global randomness versus Bell values for $N=3, 4$ at $\alpha=10$ and $\alpha=30$ under quantum attacks, respectively. The orange dashed dot line, orange real line and orange dashed line in Fig.~\ref{compare} represent min-entropy, guessing probability and optimal guessing probability of $N=3$. The gray dashed x line and gray real x line in Fig.~\ref{compare} represent min-entropy and guessing probability in Ref.~\cite{woodhead2018randomness}, respectively. The blue dashed dot line, real line and dashed line in Fig.~\ref{performance_N=4} represent min-entropy, guessing probability and optimal guessing probability of $N=4$, respectively. The results are calculated with the NPA hierarchy method at level 3.}
\end{figure}

\section{\textbf{Global Randomness versus non-maximal Bell values under quantum attacks}}
Besides certifying optimal randomness at maximal Bell values, the robustness of certifying randomness under quantum attacks at non-maximal Bell values is also of great interests for experimental researches. Specifically, the guessing probability under quantum attacks for $N$ party Bell scenario can be expressed as~\cite{woodhead2018randomness}
\begin{equation}
  G(x_1\cdots x_N|E)=\sum_{a_1\cdots a_N}p(a_1\cdots a_N(e=a_1\cdots a_N)|x_1\cdots x_N),
\end{equation}
where $E$ represents the eavesdropper's operation and $e=a_1\cdots a_N$ means successfully guessing the measurement results.

Compared with the Mermin inequality adopted in Ref.~\cite{woodhead2018randomness}, when $\alpha=10$, the guessing probability of this work is smaller at most of the non-maximal Bell value interval, which can certify more randomness (see Fig.~\ref{compare}). While, when approaching maximal Bell value, the guessing probability becomes disadvantaged as $\alpha=10$ takes finite value. In Fig.~\ref{performance_N=4}, we also present the performance of the case $N=4$ by taking $\alpha=30$. It can be expected that when taking $\alpha \rightarrow \infty$ the robutstness of certified randomness by this work can approach performing better than the Mermin inequality for the whole range of non-maximal Bell violation interval.

\section{\textbf{Bounding the Holevo quantity under quantum attacks}}
Along with certifying global randomness and local randomness in min entropy, it's also of great interests to bound the Holevo quantity $\chi(A_{N1}:E)$, which quantifies Eve's information on users' systems under quantum attacking strategies, and has various promising applications in quantum network such as quantum  conference key agreement (QCKA)~\cite{chen2007multi,zhao2020phase,grasselli2023boosting} and quantum secret sharing (QSS)~\cite{hillery1999quantum,zhang2024device},
\begin{equation}
  \chi(A_{N1}:E)=S(\rho_E)-\frac{1}{2}\sum_{a_N=0,1}S(\rho_{E|a_N}),
\end{equation}
where $S(\cdot)$ denotes the Von Neumann entropy, $\rho_E=\Tr_{A_1\cdots A_N}(|\psi_{A_1\cdots A_NE}\rangle\langle\psi_{A_1\cdots A_NE}|)$ is Eve's subsystem after tracing out users' systems. $\rho_{E|a_{N}}$ is Eve's quantum state when user $P_N$ obtained the result $a_{N}$ for the measurement $\hat{A}_{N1}$, $|\psi_{A_1\cdots A_NE}\rangle$ is Eve's purification on the state $\rho_{A_1\cdots A_N}$.

\begin{figure}
  \centering
  \vspace{0cm}
  \includegraphics[width=0.49\textwidth]{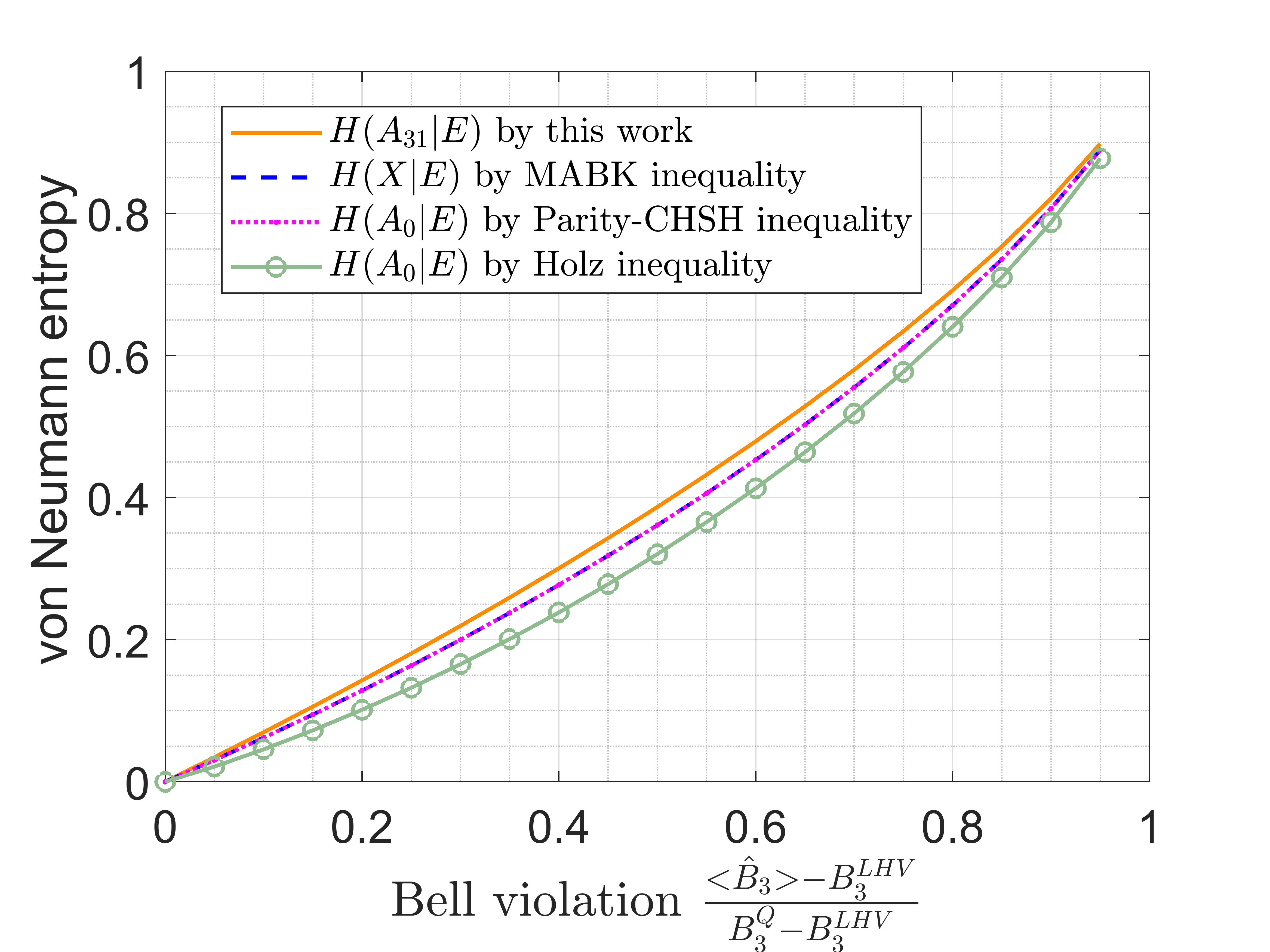}\\
  \caption{Comparison on the performance of the one-outcome conditional von Neumann entropy bounds versus Bell violation for $N=3$ parties. The orange solid line is the entropy bound by this work at $\alpha=1$. The blue dashed line is the entropy bound based on the MABK inequality~\cite{grasselli2021entropy} with $H(X|E)\geq 1-h\left(\frac{1}{2}+\frac{1}{2}\sqrt{\frac{m^2}{8}-1}\right)$ (need to note that the MABK value $m$ located beyond the genuine multipartite entanglement (GME) region, and the classical bound is substituted by GME bound $2\sqrt{2}$ here ). The pink dashed dotted line is the entropy bound based on the Parity-CHSH inequality with $H(A_0|E)\geq 1-h\left(\frac{1}{2}+\frac{1}{2}\sqrt{m^2-1}\right)$. The green circle solid line is the entropy bound based on the Holz inequality with $H(A_0|E)\geq 1-h\left(\frac{1}{4}(m+1+\sqrt{m^2+2m-3})\right)$. Here, $m$ is the Bell value for Bell inequalities correspondingly~\cite{grasselli2023boosting}.}\label{comparision:grasselli}
\end{figure}

Specifically, in order to derive analytical bounds, we consider the case $\alpha=1$. When achieving $N$-party Bell value $\langle\hat{B}_N\rangle$ for $\hat{B}_N=(\hat{A}_{10}+\hat{A}_{11})\hat{A}_{20}\cdots \hat{A}_{N0}+\sum_{i= 2}^{N}( \hat{A}_{10}-\hat{A}_{11})\hat{A}_{i1}$ with the purified quantum state $\psi_{A_1\cdots A_N E}$, the Holevo quantity after users' symmetric operations on their marginal measurement outcomes is
  \begin{equation}
    \chi(A_{N1}:E)\leq h\left(\frac{1}{2}+\frac{1}{2}\sqrt{\frac{\langle\hat{B}_N\rangle^2}{4}-(N-1)^2}\right).
  \end{equation}
  Here $h(p)=-p\log_2p-(1-p)\log_2(1-p)$ is the binary entropy. The details are expanded in the Appendix \ref{calculation_observation}. When $N=3$ parties, the result indicates the one-outcome conditional von Neumann entropy,
\begin{equation}
  H(A_{31}|E)\geq 1-h\left(\frac{1}{2}+\frac{1}{2}\sqrt{\frac{\langle\hat{B}_3\rangle^2}{4}-4}\right).
\end{equation}
In addition, we present the comparison between this work and the MABK inequality, Parity-CHSH inequality and Holz inequality in Refs.~\cite{grasselli2021entropy,holz2020genuine,ribeiro2019reply,grasselli2023boosting} at $N=3$ parties for non-maximal Bell values in Fig.~\ref{comparision:grasselli}, which indicates better performance over the current Holevo quantity bounds for promising multiparty DI cryptographic applications.

\begin{figure}
  \centering
  \vspace{0cm}
  \includegraphics[width=0.45\textwidth]{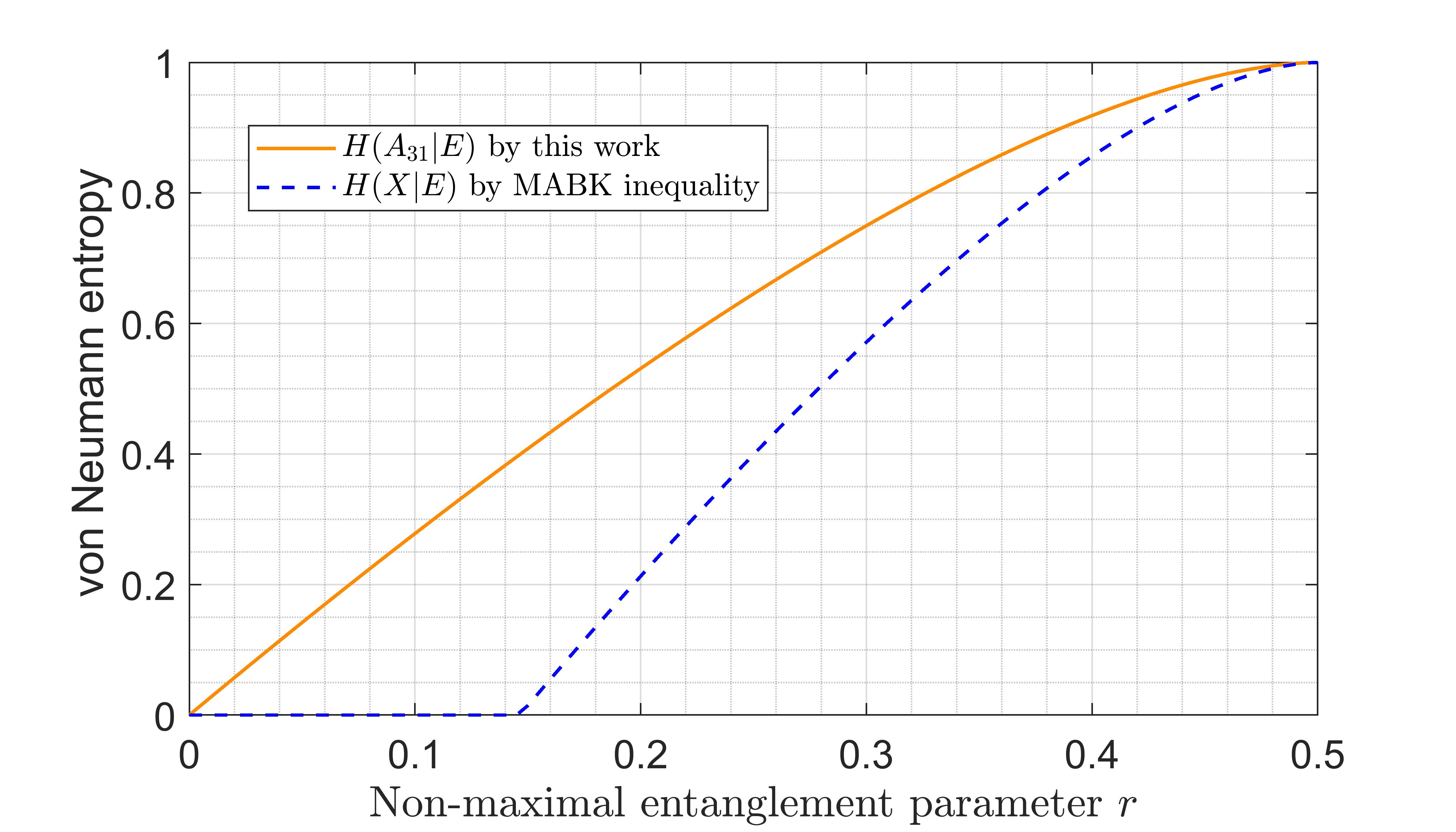}\\
  \caption{Comparison on the performance of the one-outcome conditional von Neumann entropy bounds versus non-maximal entanglement parameter $r$ for $N=3$ parties. The orange solid line is the entropy bound by this work at $\alpha=1$. The blue dashed line is the entropy bound based on the MABK inequality~\cite{grasselli2021entropy}. }\label{comparision:nonMaximal_entanglement}
\end{figure}

\section{\textbf{Performance on certifying DI quantum randomness under imperfections}} Along with the performance versus non-maximal Bell values, it is also of crucial interests to investigate the performance of randomness certification under detail kinds of imperfections from experimental perspectives.

One of the most generally investigated imperfections is the white noise for the entanglement states, i.e. Werner states. For the Werner type mixed entanglement state,
\begin{equation}
  \hat{\rho}_w^N=p|GHZ\rangle\langle GHZ|+\frac{1-p}{2^N}\hat{I},
\end{equation}
the fraction of pure GHZ state should be $p\geq \frac{B_N^{LHV}}{B_3^Q}$ to certify non-zero DI quantum randomness for the Bell inequalities constructed in this work, which is hard to have better performance compared the Mermin-type inequality in Ref.~\cite{woodhead2018randomness} especially when $N$ scales up. For example, for $N=3$ case, the global min entropy can be certified by the Mermin inequality and this work at $p_{Mermin}\geq {1/2}$ and $p\geq \frac{2}{\sqrt{5}}$ with $\alpha=1$, respectively. Meanwhile, for the one-outcome von Neumann entropy, it should be $p_{Mermin} \geq \sqrt{2}/2$ and $p \geq 2/\sqrt{5}$ to certify non-zero randomness for the Mermin inequality and this work, respectively, with $\alpha=1$ (see Appendix \ref{vonNeumannVSnonmaximal} for details).

Another generally investigated imperfections is the non-maximal entanglement states,
\begin{equation}\label{non_maximal_state}
 |\Psi\rangle=\sqrt{r}|00\cdots 0\rangle + \sqrt{1-r}|11\cdots 1\rangle
\end{equation}
where $0\leq r \leq 1$ and $r=\frac{1}{2}$ indicates the maximal GHZ state, which also performs important roles in the investigation on Bell nonlocality and related quantum information processings~\cite{white1999nonmaximally}. For example, it's been confirmed that optimal randomness can be certified via non-maximal entanglement states~\cite{acin2012randomness}. Specifically, as shown in Fig.~\ref{comparision:nonMaximal_entanglement}, when considering the pure non-maximal GHZ entanglement state in Eq.~\ref{non_maximal_state}, the constructed Bell inequalities can perform better than the MABK inequality in the task of certifying one-outcome von Neumann entropy. The reason is that non-zero von Neumann entropy can only be certified by the MABK inequality when the Bell values are located beyond the region of genuine multipartite entanglement~\cite{grasselli2021entropy}. Thus, the corresponding parameter $r$ should be increased to be approximately $r\geq 0.1465$ for the MABK inequality. While, it's $r\geq 0$ for the Bell inequalities constructed in this work to certify non-zero one-outcome von Neumann entropy when taking $\alpha=1$ (see Appendix \ref{vonNeumannVSnonmaximal} for details).

\begin{figure}
  \centering
  \vspace{0cm}
  \includegraphics[width=0.52\textwidth]{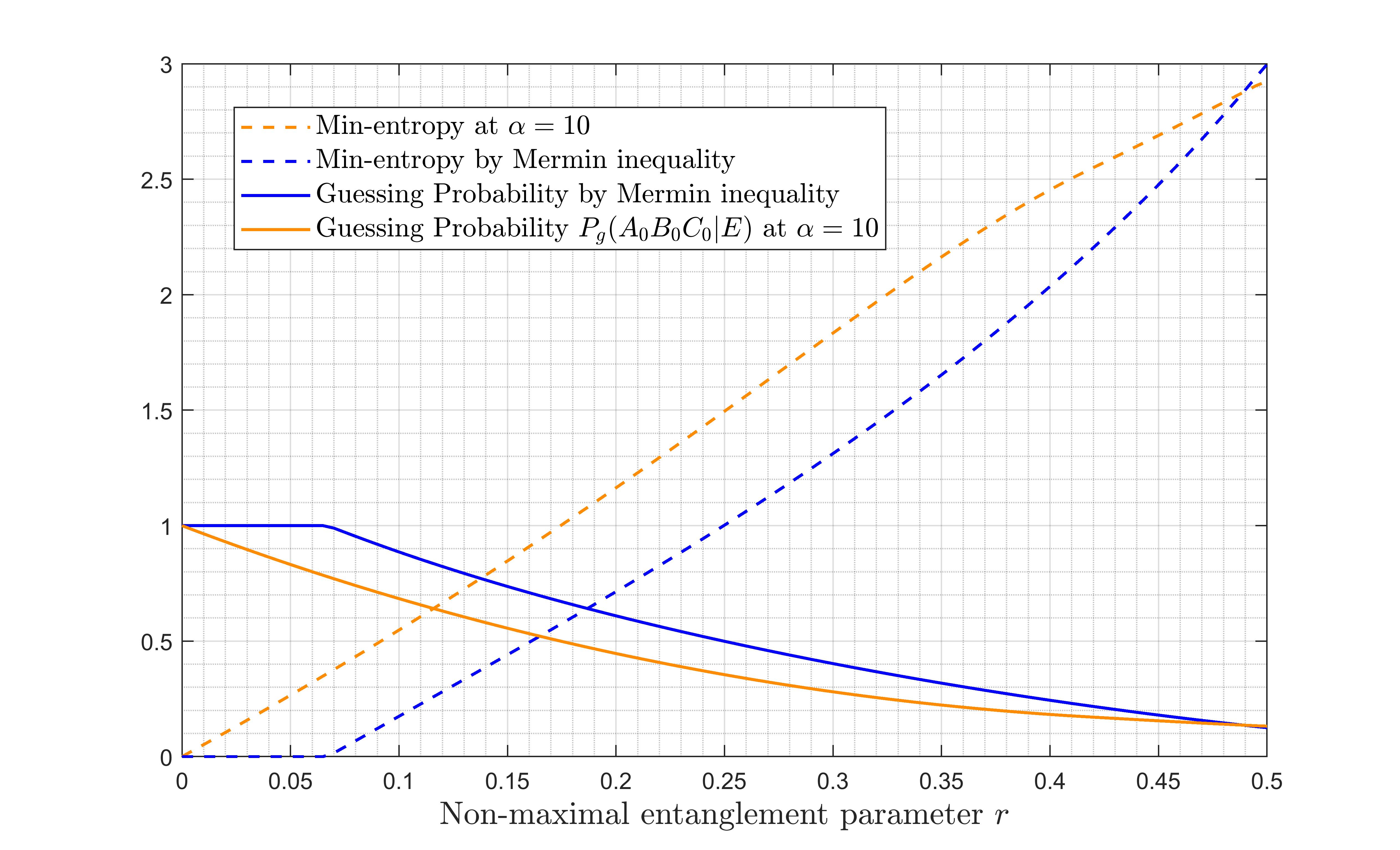}\\
  \caption{Comparison on the performance of the global min-entropy bounds versus non-maximal entanglement parameter $r$ for $N=3$ parties. The orange solid line and dashed line are the guessing probability bound and the global min-entropy bound respectively by this work at $\alpha=10$. The blue solid and dashed line are the guessing probability bound and the global min-entropy bound respectively based on the Mermin inequality. }\label{comparision:nonMaximal_entanglement_global}
\end{figure}

Besides the one-outcome von Neumann entropy, we also present the performance on the global min-entropy versus non-maximal parameter $r$. As shown in Fig.~\ref{comparision:nonMaximal_entanglement_global}, the Mermin inequality can certify non-zero global min-entropy only with approximately $r\geq 0.067$. While, it's $r\geq 0$ for the Bell inequalities constructed in this work to certify non-zero global min-entropy when taking $\alpha=10$. These two results show that the present Bell inequalities in this work can better tolerate the non-maximal parameter $r$ in the task of certifying DI quantum randomness compared with the Mermin inequality and of great promise in DI cryptographic applications.


\section{\textbf{Conclusion and Discussion}}
In this work, we have present a new family of multipartite Bell inequalities, which is constructed from the stabilizer group of the GHZ state, and can self-test GHZ states and their corresponding measurements when achieving maximal Bell values. Due to the simple representation of stabilizer group for GHZ states, this family of Bell inequalities is of simple structure and can be easily expanded to more parties when party numbers scaling up. Meanwhile, this method for constructing multipartite Bell inequalities can be applied to construct new Bell inequalities for other types of multipartite entanglement states with stabilizer groups.

Besides theoretical interests, in terms of DI randomness certification, this family of multipartite Bell inequality can certify optimal quantum randomness in min entropy with $\alpha$ increasing when achieving its maximal Bell value. Meanwhile, its robustness performance for certifying global randomness under quantum attacks at non-maximal Bell violations is also better compared with the Mermin inequality at $N=3$. Furthermore, the analytical one-outcome conditional von Neumann entropy bound is presented in this work, which also performs better compared with the MABK inequality, Parity-CHSH inequality and Holz inequality at $N=3$ parties considering non-maximal Bell violations which is crucial for DI cryptographic applications. Finally, we present the comparison on performances of the one-outcome conditional von Neumann entropy versus non-maximal entanglement GHZ state between this work and the MABK inequality, which indicates better tolerance on the non-maximal entanglement parameter $r$ of the constructed Bell inequalities in this work.

Need to remark that, as one of the most promising applications in quantum networks, the QCKA has been pushed into experimental realization for non-DI cases~\cite{pickston2023conference,yang2024experimental,Zou2026Experimental}. While, for the DIQCKA, there remains a big step. Besides the constraints on Experimental techniques, as discussed in Refs.~\cite{grasselli2023boosting,wooltorton2025genuine}, a lot of efforts still need to be involved theoretically. As the correlators in the Bell inequalities in this work are similar and simpler compared with the Holze inequality, it serves as a competitive candidate for DICKA considering the better performance on the one-outcome von Neumann entropy.

As a discussion, although an analytical tight and achievable one-outcome conditional von Neumann entropy bound is presented in this work, the strict proof remains an open problem, which is important for strict analysis on coherent quantum attacks. Meanwhile, as the reduced two party $\alpha$-CHSH inequality plays a key role in tasks like certifying quantum randomness under weak random sources and robust self-testing, it's interesting to investigate whether this family of multipartite Bell inequalities can have some advantages in certifying more quantum randomness under weak random sources and robust self-testing multipartite entanglement states.


\section{\textbf{Acknowledgments}}
S. Z. acknowledges support from the Quantum Science and Technology-National Science and Technology Major Project (Grant No.~2024ZD0302200), the Zhejiang Provincial Natural Science Foundation of China (Grant No.~LQ24A050005). R. W. acknowledges support from the Hangzhou Dianzi University start-up grant, the National Natural Science Foundation of China (Grant No.~62501220). Q. Z. acknowledges support from the Hong Kong Research Grant Council (RGC) via No. 27300823.

\begin{appendix}
\onecolumngrid

\section{Stabilizer-type Constructing Multipartite Bell inequalities}\label{counstruction_stabilizer}
For the generalized $N$-qubit GHZ state $|GHZ\rangle=\frac{1}{\sqrt{2}}(|00\cdots0\rangle+|11\cdots1\rangle)$, the corresponding generators of its stabilizer group are expressed as:
\begin{equation}\label{generator_GHZ}
\begin{split}
  \hat{g}_1&=\hat{X}_1\hat{X}_2\cdots \hat{X}_N,\\
  \hat{g}_2&=\hat{Z}_1\hat{Z}_2\hat{I}_3\cdots \hat{I}_N,\\
  \hat{g}_3&=\hat{Z}_1\hat{I}_2\hat{Z}_3\cdots \hat{I}_N,\\
  &\qquad\vdots\\
  \hat{g}_N&=\hat{Z}_1\hat{I}_2\hat{I}_3\cdots \hat{Z},
\end{split}
\end{equation}

By local transformations $\hat{A}_{10}\leftrightarrow -\hat{A}_{10}$ and $\hat{A}_{20}\leftrightarrow \hat{A}_{21}$, the $\alpha$-CHSH expression is taken as
\begin{equation}
  \hat{I}_{\alpha}'=(\hat{A}_{10}+\hat{A}_{11})\hat{A}_{20}+(\alpha \hat{A}_{10}-\hat{A}_{11})\hat{A}_{21},
\end{equation}
whose maximal Bell value can be achieved with two-qubit maximal entangled state $|\phi^{\dagger}\rangle=\frac{1}{\sqrt{2}}(|00\rangle+|11\rangle)$ for which stabilizer group is $\{\hat{X}_1\hat{X}_2,\hat{Z}_1\hat{Z}_2\}$.

For the $N$-qubit GHZ state, similarly,
inspired by Refs.~\cite{baccari2020scalable,zhao2022constructing}, taking substitutions
 \begin{equation}
 \begin{split}
   \hat{X}_1&\rightarrow \hat{A}_{10}+\hat{A}_{11},\\
   \hat{X}_2,\cdots, \hat{X}_N&\rightarrow \hat{A}_{20},\cdots,\hat{A}_{N0},\\
   \hat{Z}_1&\rightarrow \alpha\hat{A}_{10}-\hat{A}_{11},\\
   \hat{Z}_2,\cdots,\hat{Z}_N&\rightarrow \hat{A}_{21},\cdots,\hat{A}_{N1}.\\
 \end{split}
 \end{equation}
the summation for the generators in Eq.~\ref{generator_GHZ} gives the construction of this multipartite Bell inequality
\begin{equation}
  \hat{B}_{N}=(\hat{A}_{10}+\hat{A}_{11})\hat{A}_{20}\cdots \hat{A}_{N0}+\sum_{i= 2}^{N}(\alpha \hat{A}_{10}-\hat{A}_{11})\hat{A}_{i1}
\end{equation}
where \{$\hat{A}_{i0}$, $\hat{A}_{i1}$\} is the measurement setting with binary outcomes for the $i$-th party, $\alpha$ is an adjustable parameter.
Besides the GHZ state, one can also construct the multipartite for other type stabilizer states with this method.

\section{Upper bounds for the $N$-party Bell expression in quantum theory and self-testing}\label{proof_theorem}
To be more explicit, we re-present the \textbf{Theorem}~\ref{prop:1} of the main text and present its proof in this section. Without loss of generality, we only present the proof for the upper bound for quantum theory.The $N$-party Bell expression is of the following form:
\begin{equation}\label{Bell_expression}
  \hat{B}_{N}=(\hat{A}_{10}+\hat{A}_{11})\hat{A}_{20}\cdots \hat{A}_{N0}+\sum_{i= 2}^{N}(\alpha \hat{A}_{10}-\hat{A}_{11})\hat{A}_{i1},
\end{equation}
where \{$\hat{A}_{i0}$, $\hat{A}_{i1}$\} is the measurement setting with binary outcomes for the $i$-th party, $\alpha$ is an adjustable parameter. For $N\geq 3$, by linear programming, the upper bounds of the classical theory is
\begin{equation}
    B_{N}^{LHV}=\left\{
    \begin{array}{ll}
      2-(N-1)(\alpha-1), & \hbox{$\alpha\in(\alpha_L, \frac{1}{N-1}]$;} \\
      (N-1)(\alpha+1), & \hbox{$\alpha\in(\frac{1}{N-1},\infty)$.}
    \end{array}
  \right.
\end{equation}
where $\alpha_L=\frac{2N^2-2N\sqrt{N^2-2N+2}+N-1}{4N^2-5N+1}$ is obtained by the intersection of the classical and quantum bound. While, for quantum theory, the upper bound and its proof are as follows.

\begin{theorem}\label{prop:1} \textbf{Upper bound and self-testing statements}. The upper bound for the Bell expression in Eq.~\ref{Bell_expression} is
\begin{equation}
    B_{N}^{Q}=\sqrt{1+(N-1)^2\alpha^2}+\sqrt{1+(N-1)^2},
\end{equation}
which can be realized, up to local isometries, by the $N$-party GHZ state
\begin{equation}\label{ghz_state}
  |GHZ\rangle=\frac{1}{\sqrt{2}}(|00\cdots0\rangle+|11\cdots1\rangle),
\end{equation}
and the measurements
\begin{equation}~\label{obserA1}
  \begin{split}
    \hat{A}_{10}&=\frac{(N-1)\alpha}{\sqrt{1+(N-1)^2\alpha^2}}\sigma_z+\frac{1}{\sqrt{1+(N-1)^2\alpha^2}}\sigma_x,\\
    \hat{A}_{11}&=-\frac{N-1}{\sqrt{1+(N-1)^2}}\sigma_z+\frac{1}{\sqrt{1+(N-1)^2}}\sigma_x.
  \end{split}
\end{equation}
and
\begin{equation}\label{obserA2}
\begin{split}
    \hat{A}_{i0}&=\sigma_x,\\
    \hat{A}_{i1}&=\sigma_z.
  \end{split}
\end{equation}
for $2\leq i\leq N$.
\end{theorem}

\begin{proof}
The proof are presented as follows:

\begin{itemize}
  \item The quantum bound $B_{N}^{Q}=\sqrt{1+(N-1)^2\alpha^2}+\sqrt{1+(N-1)^2}$.

Here, we present the derivation of the upper bound for quantum systems when $N\geq 3$. Due to Jordan's Lemma\cite{masanes2006asymptotic,goh2018geometry}, for the Bell scenario with $N$ parties, 2 inputs and 2 outcomes, the upper bound can be achieved using qubits systems. Specifically, the observables for the $i$-th party can be modeled as
\begin{equation}\label{measurement_this}
\begin{split}
  \hat{A}_{i0}=\cos\theta_{i0}\sigma_z+\sin\theta_{i0}\sigma_x,\\
  \hat{A}_{i1}=\cos\theta_{i1}\sigma_z+\sin\theta_{i1}\sigma_x,
\end{split}
\end{equation}
where $\sigma_x$ and $\sigma_z$ are Pauli Matrices, $\theta_{i0}(\theta_{i1})$ is an adjustable angle. Therefore, the Bell value can be termed as
\begin{equation}\label{values}
  B_{N}=Tr[\hat{B}_{N}|GHZ\rangle\langle GHZ|],
\end{equation}
which is optimized when
\begin{equation}\label{angles}
  \begin{aligned}
  \tan\theta_{10}&=\frac{1}{(N-1)\alpha}, \\
  \tan\theta_{11}&=-\frac{1}{N-1},\\
  \theta_{i0}&=\pi/2, ~\text{for}~ i\geq 2,\\
  \theta_{i1}&=0, ~\text{for}~ i\geq 2,\\
  \end{aligned}
\end{equation}
which result in
\begin{equation}
  \begin{split}
    \hat{A}_{10}&=\frac{(N-1)\alpha}{\sqrt{1+(N-1)^2\alpha^2}}\sigma_z+\frac{1}{\sqrt{1+(N-1)^2\alpha^2}}\sigma_x,\\
    \hat{A}_{11}&=-\frac{N-1}{\sqrt{1+(N-1)^2}}\sigma_z+\frac{1}{\sqrt{1+(N-1)^2}}\sigma_x.
  \end{split}
\end{equation}
and
\begin{equation}
  \begin{split}
    \hat{A}_{i0}&=\sigma_x,\\
    \hat{A}_{i1}&=\sigma_z.
  \end{split}
\end{equation}
for $i\geq 2$.

\item Sum-of-square decomposition.

In above step, as the present Bell inequality is constructed from the stabilizer group of the GHZ state, we directly calculate the upper bound by optimizing the measurement settings  based on qubit systems in terms of the Jordan's lemma. This result may be doubted that the GHZ state may be not optimal, and thus the above quantum bound may be only achievable but not maximal, i.e. $B_{N}^{Q,optimal}\geq B_{N}^{Q}$. To be more complete, we present the sum-of-square (SOS) decomposition for the Bell operator $\bar{B}_N$, which can confirm that the above quantum bound is the the supremum for the constructed Bell inequality\cite{bamps2015sum}, i.e. $B_{N}^{Q,optimal}\leq B_{N}^{Q}$ once the SOS decomposition exists. Here, $\bar{B}_N$ is defined from Bell operator $\hat{B}_N$ and upper bound $B_N^Q$ as,
\begin{equation}
  \bar{B}_N=B_N^Q\hat{I}-\hat{B}_N\geq 0,
\end{equation}
which is positive semidefinite and can be proven by providing a set of operators $\{\hat{L}_j\}$ such that
\begin{equation}
  \bar{B}_N=\sum_j \hat{L}_j^\dag \hat{L}_j.
\end{equation}
where $L_j$ are polynomial functions of measurement settings $\{\hat{A}_{i0}, \hat{A}_{i1}\}$ for all $i\geq 1$. Specifically, the Polynomial operators in one of the decomposition schemes are presented as follows:
\begin{equation}
  \begin{split}
    \hat{L}_1&=\frac{1}{2b}\hat{I}-b(\hat{A}_{10}+\hat{A}_{11})\hat{A}_{20}\cdots\hat{A}_{N0},\\
    \hat{L}_2&=\frac{\sqrt{\alpha(N-1)}}{2b}\hat{I}-\frac{b}{\sqrt{\alpha(N-1)}}(\alpha\hat{A}_{10}-\hat{A}_{11})\hat{A}_{21},\\
    &\cdots\\
    \hat{L}_N&=\frac{\sqrt{\alpha(N-1)}}{2b}\hat{I}-\frac{b}{\sqrt{\alpha(N-1)}}(\alpha\hat{A}_{10}-\hat{A}_{11})\hat{A}_{N1}.
  \end{split}
\end{equation}
where $b^2=\frac{\sqrt{1+(N-1)^2\alpha^2}+\sqrt{1+(N-1)^2}\pm\sqrt{(\sqrt{1+(N-1)^2\alpha^2}+\sqrt{1+(N-1)^2})^2-\frac{(\alpha+1)^2}{\alpha}[1+\alpha(N-1)^2]}}{2\cdot (\alpha+1)^2/\alpha}$, and the operators in $\{\hat{L}_j\}$ are expanded in the basis $\{\hat{I},\hat{A}_{10}, \hat{A}_{11}\}\otimes \{\hat{I},\hat{A}_{20}, \hat{A}_{21}\}\otimes\cdots \{\hat{I},\hat{A}_{N0}, \hat{A}_{N1}\}$

As an example, for $N=3$ case, we have
\begin{equation}
  \begin{split}
    \hat{L}_1&=\frac{1}{2b}\hat{I}-b(\hat{A}_{10}+\hat{A}_{11})\hat{A}_{20}\cdots\hat{A}_{N0},\\
    \hat{L}_2&=\frac{\sqrt{2\alpha}}{2b}\hat{I}-\frac{b}{\sqrt{2\alpha}}(\alpha\hat{A}_{10}-\hat{A}_{11})\hat{A}_{21},\\
    \hat{L}_3&=\frac{\sqrt{2\alpha}}{2b}\hat{I}-\frac{b}{\sqrt{2\alpha}}(\alpha\hat{A}_{10}-\hat{A}_{11})\hat{A}_{31},
  \end{split}
\end{equation}
where $b^2=\frac{\sqrt{1+4\alpha^2}+\sqrt{5}\pm\sqrt{(\sqrt{1+4\alpha^2}+\sqrt{5})^2-\frac{(\alpha+1)^2}{\alpha}(1+4\alpha)}}{2\cdot (\alpha+1)^2/\alpha}$. Specifically, for $N=3$, we have
\begin{equation}
\begin{split}
  \bar{B}_3&=\hat{L}_1^\dag\cdot \hat{L}_1+\hat{L}_2^\dag\cdot \hat{L}_2+\hat{L}_3^\dag\cdot \hat{L}_3\\
           &=[\frac{1+4\alpha}{4b^2}+\frac{b^2(1+\alpha)^2}{\alpha}]\cdot\hat{I}-[(\hat{A}_{10}+\hat{A}_{11})\hat{A}_{20}\hat{A}_{30}+(\alpha \hat{A}_{10}-\hat{A}_{11})\hat{A}_{21}+(\alpha \hat{A}_{10}-\hat{A}_{11})\hat{A}_{31}]\\
           &=B_3^Q\hat{I}-\hat{B}_3.
\end{split}
\end{equation}
These results indicate the exist of SOS decomposition of the operators $\bar{B}_N$ and $B_{N}^{Q,optimal}\leq B_{N}^{Q}$. Thus, recalling the aforementioned $B_{N}^{Q,optimal}\geq B_{N}^{Q}$, one can confirm that the upper bound for the present multiparty Bell inequalities of quantum theory is $B_{N}^{Q}=\sqrt{1+(N-1)^2\alpha^2}+\sqrt{1+(N-1)^2}$. Meanwhile, the SOS decomposition is also an important tool for analysis on robust self-testing of entanglement states and the corresponding measurements, we leave it for future works.

\item Self-testing the GHZ state and the measurements.
%

According to above discussion, the upper bound for the $N$-party Bell inequality in \textbf{Theorem}~\ref{prop:1} can be realized by the reference GHZ state in Eq.~\ref{ghz_state} and measurements in Eq.~\ref{obserA1} and Eq.~\ref{obserA2}, up to local isometries. To self-test the GHZ state and corresponding measurements, the local isometry $\Phi=\Phi_{P_1}\otimes \Phi_{P_2}\cdots\otimes \Phi_{P_N} $  is
  \begin{equation}
    \begin{split}
      \Phi_{P_i}(|2k_i\rangle_{P_i}\otimes|0\rangle_{P_i'})& \longmapsto |2k_i\rangle_{P_i}\otimes|0\rangle_{P_i'},\\
      \Phi_{P_i}(|2k_i+1\rangle_{P_i}\otimes|0\rangle_{P_i'})&\longmapsto |2k_i\rangle_{P_i}\otimes|1\rangle_{P_i'},
    \end{split}
  \end{equation}
where $|2k_i\rangle_{P_i}$ ($|2k_i+1\rangle_{P_i}$) and $|0\rangle_{P_i'}$($|1\rangle_{P_i'}$) are computational basis for the party $P_i$'s experimental system in $k_i$-th subspace and ancillary systems~\cite{zhao2023tilted,rabelo2012device}.

The actually experimental quantum state are high dimensional and can be expression as
\begin{equation}
|\tilde{GHZ}\rangle=\oplus_{k_1,k_2,\cdots,k_N}\sqrt{q_{k_1,k_2,\cdots,k_N}}|GHZ\rangle_{k_1,k_2,\cdots,k_N}
\end{equation}
with $|GHZ\rangle_{k_1,k_2,\cdots,k_N} =\frac{1}{\sqrt{2}}(|2k_1,2k_2,\cdots,2k_N\rangle+|2k_1+1,2k_2+1,\cdots,2k_N+1\rangle)$, $q_{k_1,k_2,\cdots,k_N}\geq 0$ and $\sum q_{k_1,k_2,\cdots,k_N}=1 $. Meanwhile, the measurement observables are
\begin{equation}
  \begin{split}
    \tilde{\hat{A}}_{i0}&=\oplus_{k_i}\hat{A}_{i0}^{k_i} , \\
    \tilde{\hat{A}}_{i1}&=\oplus_{k_i}\hat{A}_{i1}^{k_i}.
  \end{split}
\end{equation}
with
\begin{equation}
  \begin{split}
    \hat{A}_{10}^{k_1}&=\frac{(N-1)\alpha}{\sqrt{1+(N-1)^2\alpha^2}}(|2k_1\rangle\langle2k_1|-|2k_1+1\rangle\langle2k_1+1|)+\frac{1}{\sqrt{1+(N-1)^2\alpha^2}}(|2k_1\rangle\langle2k_1+1|+|2k_1+1\rangle\langle2k_1|),\\
    \hat{A}_{11}^{k_1}&=-\frac{N-1}{\sqrt{1+(N-1)^2}}(|2k_1\rangle\langle2k_1|-|2k_1+1\rangle\langle2k_1+1|)+\frac{1}{\sqrt{1+(N-1)^2}}(|2k_1\rangle\langle2k_1+1|+|2k_1+1\rangle\langle2k_1|),
  \end{split}
\end{equation}

and

\begin{equation}
  \begin{split}
    \hat{A}_{i0}^{k_i}&=|2k_i\rangle\langle2k_i+1|+|2k_i+1\rangle\langle2k_i|,\\
    \hat{A}_{i1}^{k_i}&=|2k_i\rangle\langle2k_i|-|2k_i+1\rangle\langle2k_i+1|,
  \end{split}
\end{equation}
for $2\leq i\leq N$.
Thus, the reference state $|GHZ\rangle$ and measurements $\hat{A}_{i0}$ and $\hat{A}_{i1}$ can be self-tested from the experimental state $|\tilde{GHZ}\rangle$ and $\tilde{\hat{A}}_{i0}$ and $\tilde{\hat{A}}_{i1}$ as follows:
\begin{equation}
  \begin{split}
    &\Phi(|\tilde{GHZ}\rangle\otimes |00\cdots 0\rangle_{P_1'P_2'\cdots P_N'})\\
    &=\Phi(\oplus_{k_1,k_2,\cdots,k_N}\sqrt{q_{k_1,k_2,\cdots,k_N}}\frac{1}{\sqrt{2}}(|2k_1,2k_2,\cdots,2k_N\rangle+|2k_1+1,2k_2+1,\cdots,2k_N+1\rangle)|00\cdots 0\rangle_{P_1'P_2'\cdots P_N'})\\
    &=\oplus_{k_1,k_2,\cdots,k_N}\sqrt{q_{k_1,k_2,\cdots,k_N}}\frac{1}{\sqrt{2}}(|2k_1,2k_2,\cdots,2k_N\rangle|00\cdots 0\rangle_{P_1'P_2'\cdots P_N'}+|2k_1,2k_2,\cdots,2k_N\rangle|11\cdots 1\rangle_{P_1'P_2'\cdots P_N'})\\
    &=\oplus_{k_1,k_2,\cdots,k_N}\sqrt{q_{k_1,k_2,\cdots,k_N}}\frac{1}{\sqrt{2}}|2k_1,2k_2,\cdots,2k_N\rangle\frac{1}{\sqrt{2}}(|00\cdots 0\rangle_{P_1'P_2'\cdots P_N'}+|11\cdots 1\rangle_{P_1'P_2'\cdots P_N'})\\
    &=|junk\rangle_{P_1P_2\cdots P_N}|GHZ\rangle_{P_1'P_2'\cdots P_N'}.
  \end{split}
\end{equation}
and

\begin{equation}
  \begin{split}
     &\Phi(\tilde{\hat{A}}_{10}|\tilde{GHZ}\rangle\otimes |00\cdots 0\rangle_{P_1'P_2'\cdots P_N'})\\
     &=\Phi(\oplus_{k_1,k_2,\cdots,k_N}\sqrt{q_{k_1,k_2,\cdots,k_N}}(\frac{(N-1)\alpha}{\sqrt{1+(N-1)^2\alpha^2}}(|2k_1\rangle\langle2k_1|-|2k_1+1\rangle\langle2k_1+1|)\\
     &+\frac{1}{\sqrt{1+(N-1)^2\alpha^2}}(|2k_1\rangle\langle2k_1+1|+|2k_1+1\rangle\langle2k_1|))\frac{1}{\sqrt{2}}(|2k_1,2k_2,\cdots,2k_N\rangle\\
     &+|2k_1+1,2k_2+1,\cdots,2k_N+1\rangle)|00\cdots 0\rangle_{P_1'P_2'\cdots P_N'}))\\
     &=\Phi( \oplus_{k_1,k_2,\cdots,k_N}\frac{\sqrt{q_{k_1,k_2,\cdots,k_N}}}{\sqrt{2}}(\frac{(N-1)\alpha}{\sqrt{1+(N-1)^2\alpha^2}}(|2k_1,2k_2,\cdots,2k_N\rangle-|2k_1+1,2k_2+1,\cdots,2k_N+1\rangle)\\
     &+\frac{1}{\sqrt{1+(N-1)^2\alpha^2}}(|2k_1,2k_2+1,\cdots,2k_N+1\rangle+|2k_1+1,2k_2,\cdots,2k_N\rangle))|00\cdots 0\rangle_{P_1'P_2'\cdots P_N'} )\\
     &=\oplus_{k_1,k_2,\cdots,k_N}\frac{\sqrt{q_{k_1,k_2,\cdots,k_N}}}{\sqrt{2}}(\frac{(N-1)\alpha}{\sqrt{1+(N-1)^2\alpha^2}}(|2k_1,2k_2,\cdots,2k_N\rangle|00\cdots 0\rangle_{P_1'P_2'\cdots P_N'}+|2k_1,2k_2,\cdots,2k_N\rangle|11\cdots 1\rangle_{P_1'P_2'\cdots P_N'})\\
     &+\frac{1}{\sqrt{1+(N-1)^2\alpha^2}}(|2k_1,2k_2,\cdots,2k_N\rangle|01\cdots 1\rangle_{P_1'P_2'\cdots P_N'}+|2k_1,2k_2,\cdots,2k_N\rangle|10\cdots 0\rangle_{P_1'P_2'\cdots P_N'}))\\
     &=|junk\rangle_{P_1P_2\cdots P_N}\frac{1}{\sqrt{2}}( \frac{(N-1)\alpha}{\sqrt{1+(N-1)^2\alpha^2}}(|00\cdots 0\rangle-|11\cdots 1\rangle)+\frac{1}{\sqrt{1+(N-1)^2\alpha^2}}(|01\cdots 1\rangle+|10\cdots 0\rangle))_{P_1'P_2'\cdots P_N'}\\
     &=|junk\rangle_{P_1P_2\cdots P_N} \hat{A}_{10}|GHZ\rangle_{P_1'P_2'\cdots P_N'}.
  \end{split}
\end{equation}
and
\begin{equation}
  \begin{split}
     &\Phi(\tilde{\hat{A}}_{11}|\tilde{GHZ}\rangle\otimes |00\cdots 0\rangle_{P_1'P_2'\cdots P_N'})\\
     &=\Phi(\oplus_{k_1,k_2,\cdots,k_N}\sqrt{q_{k_1,k_2,\cdots,k_N}}(-\frac{N-1}{\sqrt{1+(N-1)^2}}(|2k_1\rangle\langle2k_1|-|2k_1+1\rangle\langle2k_1+1|)\\
     &+\frac{1}{\sqrt{1+(N-1)^2}}(|2k_1\rangle\langle2k_1+1|+|2k_1+1\rangle\langle2k_1|))\frac{1}{\sqrt{2}}(|2k_1,2k_2,\cdots,2k_N\rangle\\
     &+|2k_1+1,2k_2+1,\cdots,2k_N+1\rangle)|00\cdots 0\rangle_{P_1'P_2'\cdots P_N'}))\\
     &=\Phi( \oplus_{k_1,k_2,\cdots,k_N}\frac{\sqrt{q_{k_1,k_2,\cdots,k_N}}}{\sqrt{2}}(-\frac{(N-1)}{\sqrt{1+(N-1)^2}}(|2k_1,2k_2,\cdots,2k_N\rangle-|2k_1+1,2k_2+1,\cdots,2k_N+1\rangle)\\
     &+\frac{1}{\sqrt{1+(N-1)^2}}(|2k_1,2k_2+1,\cdots,2k_N+1\rangle+|2k_1+1,2k_2,\cdots,2k_N\rangle))|00\cdots 0\rangle_{P_1'P_2'\cdots P_N'} )\\
     &=\oplus_{k_1,k_2,\cdots,k_N}\frac{\sqrt{q_{k_1,k_2,\cdots,k_N}}}{\sqrt{2}}(-\frac{(N-1)}{\sqrt{1+(N-1)^2}}(|2k_1,2k_2,\cdots,2k_N\rangle|00\cdots 0\rangle_{P_1'P_2'\cdots P_N'}+|2k_1,2k_2,\cdots,2k_N\rangle|11\cdots 1\rangle_{P_1'P_2'\cdots P_N'})\\
     &+\frac{1}{\sqrt{1+(N-1)^2}}(|2k_1,2k_2,\cdots,2k_N\rangle|01\cdots 1\rangle_{P_1'P_2'\cdots P_N'}+|2k_1,2k_2,\cdots,2k_N\rangle|10\cdots 0\rangle_{P_1'P_2'\cdots P_N'}))\\
     &=|junk\rangle_{P_1P_2\cdots P_N}\frac{1}{\sqrt{2}}( -\frac{(N-1)}{\sqrt{1+(N-1)^2}}(|00\cdots 0\rangle-|11\cdots 1\rangle)+\frac{1}{\sqrt{1+(N-1)^2}}(|01\cdots 1\rangle+|10\cdots 0\rangle))_{P_1'P_2'\cdots P_N'}\\
     &=|junk\rangle_{P_1P_2\cdots P_N} \hat{A}_{11}|GHZ\rangle_{P_1'P_2'\cdots P_N'}.
  \end{split}
\end{equation}
where $|junk\rangle_{P_1P_2\cdots P_N}$ is a junk state.

Similarly, the self-testing of measurement observables $\hat{A}_{i0}$ and $\hat{A}_{i1}$ ($2\leq i\leq N$) can be obtained.

\end{itemize}
\end{proof}

\section{Calculating upper bound on the Holevo quantity when taking $\alpha=1$ for multiparty device-independent quantum cryptographic applications}\label{calculation_observation}
Considering the multiple rounds experiments described in the Fig.1 of the main text for the $(N, 2, 2)$ Bell scenarios, each party $P_i$ ($i\in\{1,2,\cdots,N\}$) makes measurement choices with $x_i\in\{\hat{A}_{i0}, \hat{A}_{i1}\}$ and obtains outcomes with $a_i\in\{0,1\}$. The measurement results can be summarized as a probability distribution $p(a_1a_2\cdots a_N|x_1x_2\cdots x_N)$, from which the Bell value for the $N$-party Bell inequality is
\begin{equation}
  \langle \hat{B}_N\rangle=\langle(\hat{A}_{10}+\hat{A}_{11})\hat{A}_{20}\cdots \hat{A}_{N0}\rangle+\sum_{i= 2}^{N}\langle( \hat{A}_{10}-\hat{A}_{11})\hat{A}_{i1}\rangle.
\end{equation}
Here in this part, we have slightly abused the notation $\hat{B}_N$ without causing misunderstandings.

Specifically, the Bell value quantifies the nonlocal correlation among users, and can be used to indicated the correlation between users and Eve's systems. Here we would like to put out the analytical relation of the Holevo quantity $ \chi(A_{N1}:E)$ versus the observed Bell value $\langle \hat{B}_N \rangle$ with
\begin{equation}\label{holevo}
  \chi(A_{N1}:E)=S(\rho_E)-\frac{1}{2}\sum_{a_N=0,1}S(\rho_{E|a_N}),
\end{equation}
where $S(\cdot)$ denotes the Von Neumann entropy, $\rho_E=\Tr_{A_1\cdots A_N}(|\psi_{A_1\cdots A_NE}\rangle\langle\psi_{A_1\cdots A_NE}|)$ is Eve's subsystem after tracing out users' systems. $\rho_{E|a_{N1}}$ is Eve's quantum state when user $P_N$ obtained the result $a_{N1}$ for the measurement $\hat{A}_{N1}$, $|\psi_{A_1\cdots A_NE}\rangle$ is Eve's purification on the state $\rho_{A_1\cdots A_N}$. Specifically, we consider the case $\alpha=1$ due to the symmetry for easily analytical calculation, the results are presented as follows. While, for other cases that $\alpha\neq 1$, one can obtain the bounds by numerical methods and we leave it as an open problem.

\begin{observation}\label{prop:2}
  When achieving $N$-party Bell value $\langle\hat{B}_N\rangle$ for $\hat{B}_N=(\hat{A}_{10}+\hat{A}_{11})\hat{A}_{20}\cdots \hat{A}_{N0}+\sum_{i= 2}^{N}( \hat{A}_{10}-\hat{A}_{11})\hat{A}_{i1}$ with the purified quantum state $\psi_{A_1\cdots A_N E}$, the Holevo quantity after users' symmetric operations on their marginal measurement outcomes is
  \begin{equation}
    \chi(A_{N1}:E)\leq h\left(\frac{1}{2}+\frac{1}{2}\sqrt{\frac{\langle\hat{B}_N\rangle^2}{4}-(N-1)^2}\right).
  \end{equation}
  Here $h(p)=-p\log_2p-(1-p)\log_2(1-p)$ is the binary entropy.
\end{observation}

The calculation on the upper bound of the Holevo quantity is inspired by Refs.~\cite{pironio2009device,grasselli2021entropy} with following steps.
\begin{description}
  \item[Step 1]  \emph{Reduction to two dimensional quantum systems for the $(N,2,2)$ Bell scenarios.} It has been well established that due to Jordan's Lemma\cite{masanes2006asymptotic,goh2018geometry}, for the Bell scenario with $N$ parties, 2 inputs and 2 outcomes, the upper bound for the multipatite Bell value can be achieved using qubits systems~\cite{pironio2009device,grasselli2021entropy}. Here we directly use this conclusion without detail proofs.

       \begin{table}[h]
  \caption{Transformation results under operators related within the Bell expression $\hat{B}_3$.}\label{table:transformation}
      \begin{tabular*}{0.876\textwidth}{|c|c|c|c|c|c|c|c|c|c|}
        \hline
         & $\sigma_x\otimes\sigma_x\otimes\sigma_x$ & $\sigma_x\otimes\sigma_x\otimes\sigma_z$ & $\sigma_x\otimes\sigma_z\otimes\sigma_x$ & $\sigma_x\otimes\sigma_z\otimes\sigma_z$ & $\sigma_z\otimes\sigma_x\otimes\hat{I}$ & $\sigma_z\otimes\sigma_z\otimes\hat{I}$ & $\sigma_z\otimes\hat{I}\otimes\sigma_x$ &$\sigma_z\otimes\hat{I}\otimes\sigma_z$ & classification \\
         \hline
        $\psi_{000}$ & $\psi_{000}$ & $-\psi_{101}$ & $-\psi_{110}$ & $\psi_{011}$ & $\psi_{110}$ & $\psi_{000}$ & $\psi_{101}$ &$\psi_{000}$ & class 1\\
        \hline
        $\psi_{100}$ & $-\psi_{100}$ & $\psi_{001}$ & $\psi_{010}$ & $-\psi_{111}$ & $\psi_{010}$ & $\psi_{100}$ & $\psi_{001}$ & $\psi_{100}$ & class 2 \\
        \hline
        $\psi_{001}$ & $\psi_{001}$ & $\psi_{100}$ & $-\psi_{111}$ & $-\psi_{010}$ & $\psi_{111}$ & $\psi_{001}$ & $\psi_{100}$ & $-\psi_{010}$ & class 2\\
        \hline
        $\psi_{101}$ & $-\psi_{101}$ & $-\psi_{000}$ & $\psi_{011}$ & $\psi_{110}$ & $\psi_{011}$ & $\psi_{101}$ & $\psi_{000}$ & $\psi_{101}$ & class 1\\
        \hline
        $\psi_{010}$ & $\psi_{010}$ & $\psi_{111}$ & $\psi_{100}$ & $-\psi_{001}$ & $\psi_{100}$ & $-\psi_{010}$ & $\psi_{111}$ & $\psi_{010}$ & class 2\\
        \hline
        $\psi_{110}$ & $-\psi_{110}$ & $\psi_{011}$ & $-\psi_{000}$ & $-\psi_{101}$ & $\psi_{000}$ & $-\psi_{110}$ & $\psi_{011}$ & $\psi_{110}$ & class 1\\
        \hline
        $\psi_{011}$ & $\psi_{011}$ & $\psi_{110}$ & $\psi_{101}$ & $\psi_{000}$ & $\psi_{101}$ & $-\psi_{011}$ & $\psi_{110}$ & $-\psi_{011}$ & class 1\\
        \hline
        $\psi_{111}$ & $-\psi_{111}$ & $-\psi_{010}$ & $-\psi_{001}$ & $-\psi_{100}$ & $\psi_{001}$ & $-\psi_{111}$ & $\psi_{010}$ & $-\psi_{111}$ & class 2\\
        \hline
      \end{tabular*}
      \end{table}
  \item[Step 2] \emph{Reduction in the GHZ basis.} We firstly start from $N=3$ cases, for which the three-qubit GHZ basis vectors are:
      \begin{equation}
        |\psi_{ijk}\rangle=\frac{1}{\sqrt{2}}(|0~j~k\rangle+(-1)^i|1~\bar{j}~\bar{k}\rangle),~\forall i,j,k \in\{0,1\}.
      \end{equation}
      We consider that the entanglement source is controlled by Eve who is allowed to have quantum registers and holds purifications on users states. Without loss of generality, we calculate the Holevo quantity in $X-Z$ plane. The observations can be modeled as
      \begin{equation}\label{measurement_xz}
          \begin{split}
            \hat{A}_{10}&=\cos\alpha\sigma_z+\sin\alpha\sigma_x,\\
            \hat{A}_{11}&=-\cos\alpha\sigma_z+\sin\alpha\sigma_x.\\
            \hat{A}_{20}&=\cos\beta\sigma_z+\sin\beta\sigma_x,\\
            \hat{A}_{21}&=\cos\beta\sigma_z-\sin\beta\sigma_x.\\
            \hat{A}_{30}&=\cos\gamma\sigma_z+\sin\gamma\sigma_x,\\
            \hat{A}_{31}&=\cos\gamma\sigma_z-\sin\gamma\sigma_x.
          \end{split}
        \end{equation}
      The Bell expression is described as
      \begin{equation}
      \begin{split}
        \hat{B}_3=&2\sin\alpha\sigma_x\otimes(\cos\beta\sigma_z+\sin\beta\sigma_x)\otimes(\cos\gamma\sigma_z+\sin\gamma\sigma_x)+2\cos\alpha\sigma_z\otimes(\cos\beta\sigma_z-\sin\beta\sigma_x)\otimes\hat{I}\\
        &+2\cos\alpha\sigma_z\otimes\hat{I}\otimes(\cos\gamma\sigma_z-\sin\gamma\sigma_x),
      \end{split}
      \end{equation}
      in which the related operators are
      \begin{equation}
        \{\sigma_x\otimes\sigma_x\otimes\sigma_x,\sigma_x\otimes\sigma_x\otimes\sigma_z,\sigma_x\otimes\sigma_z\otimes\sigma_x,\sigma_x\otimes\sigma_z\otimes\sigma_z,\sigma_z\otimes\sigma_x\otimes\hat{I},\sigma_z\otimes\sigma_z\otimes\hat{I},\sigma_z\otimes\hat{I}\otimes\sigma_x,\sigma_z\otimes\hat{I}\otimes\sigma_z\}.
      \end{equation}

      According to Table~\ref{table:transformation}, the GHZ basis vectors under the transformations related with the Bell expression $\hat{B}_3$ are classified into two classes
      \begin{equation}
        \{|\psi_{000}\rangle,|\psi_{101}\rangle,|\psi_{110}\rangle,|\psi_{011}\rangle\} \quad \text{and}\quad  \{|\psi_{100}\rangle,|\psi_{001}\rangle,|\psi_{010}\rangle,|\psi_{111}\rangle\}
      \end{equation}
      Thus, a general quantum states can be described as the mix of pure states in above two classes
      \begin{equation}
        \rho=\lambda_1\rho_1+\lambda_2\rho_2
      \end{equation}
      with $\rho_1$ and $\rho_2$ are the density matrix for $\psi_1=\sqrt{\lambda_{000}}\psi_{000}+\sqrt{\lambda_{101}}e^{-i \phi_{101}}\psi_{101}+\sqrt{\lambda_{110}}e^{-i \phi_{110}}\psi_{110}+\sqrt{\lambda_{011}}e^{-i \phi_{011}}\psi_{011}$ and $\psi_2=\sqrt{\lambda_{100}}\psi_{100}+\sqrt{\lambda_{001}}e^{-i \phi_{001}}\psi_{001}+\sqrt{\lambda_{010}}e^{-i \phi_{010}}\psi_{010}+\sqrt{\lambda_{111}}e^{-i \phi_{111}}\psi_{111}$, respectively. Here, $1\geq \lambda_1,\lambda_2\geq 0$ and $\lambda_1+\lambda_2=1$, $\lambda_{000}+\lambda_{101}+\lambda_{110}+\lambda_{011}=1$ and $\lambda_{100}+\lambda_{001}+\lambda_{010}+\lambda_{111}=1$.

      Need to note that the upper bound for the Bell expression can be saturated by the following four mixed states.
  \begin{equation}\label{saturate_states}
    \begin{split}
      \rho_{\lambda_{000}}=\lambda_{000}|\psi_{000}\rangle\langle\psi_{000}|+(1-\lambda_{000})|\psi_{100}\rangle\langle\psi_{100}|,\\
      \rho_{\lambda_{001}}=\lambda_{001}|\psi_{001}\rangle\langle\psi_{001}|+(1-\lambda_{001})|\psi_{101}\rangle\langle\psi_{101}|,\\
      \rho_{\lambda_{010}}=\lambda_{010}|\psi_{010}\rangle\langle\psi_{010}|+(1-\lambda_{010})|\psi_{110}\rangle\langle\psi_{110}|,\\
      \rho_{\lambda_{011}}=\lambda_{011}|\psi_{011}\rangle\langle\psi_{011}|+(1-\lambda_{011})|\psi_{111}\rangle\langle\psi_{111}|,\\
    \end{split}
  \end{equation}

    In the following, we will calculate the Holevo quantity from one of the quantum stated saturating the Bell expression.

     \item[Step 3] \emph{Upper bound for three party Bell expression.}

    The Bell value $\langle\hat{B}_3\rangle$ is calculated with Horodecki's theorem \cite{horodecki1995violating}. Specifically, a general $N$-qubit mix state can be described in Pauli basis with
      \begin{equation}\label{correlationM1}
        \rho=\frac{1}{2^N}\sum_{l_1,l_2,\cdots,l_N\in\{0,1,2,3\}}\tau_{l_1l_2l_3}\sigma_{l_1}\otimes\sigma_{l_2}\otimes\cdots\otimes\sigma_{l_N},
      \end{equation}
      with $\tau_{l_1l_2\cdots l_N}=\Tr(\rho\sigma_{l1}\otimes\sigma_{l_2}\otimes\cdots\otimes\sigma_{l_N})$, $\sigma_1,\sigma_2,\sigma_3$ are Pauli matrix $\sigma_x,\sigma_y,\sigma_z$ respectively, and $\sigma_{0}$ is identity matrix. From Eq.\ref{correlationM1}, one can easily obtain the correlation matrix $T_{\rho}$ with elements $ [T_{\rho}]_{mn}=\Tr[\rho\sigma_{l_1}\otimes\sigma_{l_2}\otimes\cdots\otimes\sigma_{l_N}]$, such that
      \begin{equation}\label{correlationIndex}
      \begin{split}
        m=1+\sum_{k=1}^{\lfloor N/2\rfloor}3^{\lfloor N/2\rfloor-k}(l_k-1),  \\
        n=1+\sum_{k=\lfloor N/2\rfloor+1}^N 3^{N-k}(l_k-1),
      \end{split}
      \end{equation}
      where $l_1,l_2,\cdots,l_N\in{1,2,3}$, $\lfloor\cdot\rfloor$ is the floor function giving the greatest integer smaller or equal to the input. For one of the quantum state saturating Bell expression in Eq.~\ref{saturate_states}, $\rho_{\lambda_{000}}$, we have
      \begin{equation}\label{correlationM2}
      T_{\rho_{\lambda}}=
        \left(
                             \begin{array}{ccccccccc}
                               \lambda_{000}-\lambda_{100} & 0 & 0 & 0 & -\lambda_{000}+\lambda_{100} & 0 & 0 & 0& 0 \\
                               0 & -\lambda_{000}+\lambda_{100}& 0 & -\lambda_{000}+\lambda_{100}& 0 & 0 & 0 & 0 & 0 \\
                               0 & 0 & 0 & 0 & 0 & 0 & 0 & 0 & 0 \\
                             \end{array}
                           \right).
      \end{equation}
      where $\lambda_{100}=1-\lambda_{000}$

      Recalling that $3$-party Bell inequality is $\hat{B}_3=(\hat{A}_{10}+\hat{A}_{11})\hat{A}_{20}\hat{A}_{30}+( \hat{A}_{10}-\hat{A}_{11})\hat{A}_{21}+( \hat{A}_{10}-\hat{A}_{11})\hat{A}_{31}$, the correlation matrices $T'_{\rho_{\lambda}}$ and $T''_{\rho_{\lambda}}$ for the reduced density matrices by tracing out user $P_3$ and $P_2$'s systems respectively are also crucial,
      \begin{equation}\label{correlationM3}
        T'_{\rho_{\lambda}}=\left(
                              \begin{array}{ccc}
                                0 & 0 & 0 \\
                                0 & 0 & 0 \\
                                0 & 0 & \lambda_{000}+\lambda_{100} \\
                              \end{array}
                            \right)
      \end{equation}
     and
       \begin{equation}\label{correlationM4}
        T''_{\rho_{\lambda}}=\left(
                              \begin{array}{ccc}
                                0 & 0 & 0 \\
                                0 & 0 & 0 \\
                                0 & 0 & \lambda_{000}+\lambda_{100} \\
                              \end{array}
                            \right).
      \end{equation}
      Thus, the upper bound of Bell value for give quantum states can be expressed as
      \begin{equation}\label{optimization}
      \begin{split}
        \langle \hat{B}_{3,\rho_\lambda}\rangle&=\max_{\vec{a_{i0}},\vec{a_{i1}},\forall i\in{1,2,3}} (\vec{a_{10}}+\vec{a_{11}})T_{\rho_\lambda}\vec{a_{20}}\vec{a_{30}}+ (\vec{a_{10}}-\vec{a_{11}})T'_{\rho_{\lambda}}\vec{a_{21}}+(\vec{a_{10}}-\vec{a_{11}})T''_{\rho_{\lambda}}\vec{a_{31}}\\
        &=\max_{\theta,\vec{a'_{10}},\vec{a'_{11}},\vec{a_{i0}},\vec{a_{i1}},\forall i\in{2,3}} 2\cos\theta\vec{a'_{10}}T_{\rho_\lambda}\vec{a_{20}}\vec{a_{30}}+2\sin\theta\vec{a'_{11}}T'_{\rho_\lambda}\vec{a_{21}}+2\sin\theta\vec{a'_{11}}T''_{\rho_\lambda}\vec{a_{31}}\\
        &=\max_{\theta} 2(\cos\theta t_0+\sin\theta t_1 +\sin\theta t_2),
      \end{split}
      \end{equation}
      where $\vec{a_{i0}},\vec{a_{i1}},\forall i\in{1,2,3}$ are normalized three dimensional vectors for users' observables. The second equation is obtained by taking $\vec{a_{10}}+\vec{a_{11}}=2\cos\theta\vec{a'_{10}}$, $\vec{a_{10}}-\vec{a_{11}}=2\sin\theta\vec{a'_{11}}$, $\vec{a'_{10}}\perp\vec{a'_{11}}$ and $||\vec{a'_{10}}||=||\vec{a'_{11}}||=1$.

      From Eqs.~\ref{correlationM2}, \ref{correlationM3}, \ref{correlationM4}, it can be easily derived that
      \begin{equation}
        \begin{split}
          t_0=&|\lambda_{000}-\lambda_{100}|,\\
          t_1=&|\lambda_{000}+\lambda_{100}|=1,\\
          t_2=&|\lambda_{000}+\lambda_{100}|=1.
        \end{split}
      \end{equation}

  For the state $\rho(\lambda_{000})$, one can obtain that
  \begin{equation}\label{bellentropy}
  \begin{split}
    \langle \hat{B}_{3,\rho_{\lambda_{000}}}\rangle&=\max_{\theta}2\left(\cos\theta(2\lambda_{000}-1)+2\sin\theta\right)\\
    &=2\sqrt{(2\lambda_{000}-1)^2+4}.
  \end{split}
  \end{equation}
  which is obtained by $\frac{\cos\theta}{\sin\theta}=\frac{2\lambda_{000}-1}{2}$.
  When taking $\lambda_{000}=1$, it's $\langle \hat{B}_{3,\rho_{\lambda_{000}}}\rangle=2\sqrt{5}$, which coincides with the results in \textbf{Theorem}~\ref{prop:1} with $\alpha=1$. Meanwhile, there are similar results for other states in Eq.~\ref{saturate_states}.

  \item[Step 4] \emph{Calculation on the Holevo quantity}.
  From the quantum state $\rho_{\lambda_{000}}=\lambda_{000}|\psi_{000}\rangle\langle\psi_{000}|+(1-\lambda_{000})|\psi_{000}\rangle\langle\psi_{000}|$ in Eq.~\ref{saturate_states}, the Eve's purified quantum state is
  \begin{equation}
    |\phi_{000}\rangle=\sqrt{\lambda_{000}}|\psi_{000}\rangle|e_0\rangle+\sqrt{1-\lambda_{000}}|\psi_{100}\rangle|e_1\rangle,
  \end{equation}
  where $\{|e_0\rangle, |e_1\rangle\}$ is the orthogonal basis for Eve's system. By tracing out users $P_1$ and $P_2$'s systems, the reduced density matrix is
  \begin{equation}\label{p3E}
  \begin{split}
    \rho^{P_3,E}_{\phi_{000}}=&\frac{\lambda_{000}}{2}(|0\rangle\langle 0|+|1\rangle\langle 1|)|e_0\rangle\langle e_0|+\frac{\sqrt{\lambda_{000}(1-\lambda_{000})}}{2}(|0\rangle\langle 0|-|1\rangle\langle 1|)|e_0\rangle\langle e_1|\\
    &+\frac{\sqrt{\lambda_{000}(1-\lambda_{000})}}{2}(|0\rangle\langle 0|-|1\rangle\langle 1|)|e_1\rangle\langle e_0|+\frac{1-\lambda_{000}}{2}(|0\rangle\langle 0|+|1\rangle\langle 1|)|e_1\rangle\langle e_1|.
  \end{split}
  \end{equation}
  Supposing that user $P_3$'s projective measurement is $\{|c_1^0\rangle\langle c_1^0|,|c_1^1\rangle\langle c_1^1|\}$ which related with his observable $\hat{A}_{31}$ in the $X-Z$ plane with
  \begin{equation}\label{basisvec}
  \begin{split}
    |c_1^0\rangle&=\cos\theta'|0\rangle+\sin\theta'|1\rangle,\\
    |c_1^1\rangle&=\sin\theta'|0\rangle-\cos\theta'|1\rangle,
  \end{split}
  \end{equation}
  the Eve's normalized states after user $P_3$ obtains outcomes $a_3=0$ and $a_3=1$ by measuring $\rho^{P_3,E}_{\phi_{000}}$ are
  \begin{equation}
    \begin{split}
      \rho^{E}_{\phi_{000}|c_1^0}=\lambda_{000}|e_0\rangle\langle e_0|+\cos2\theta'|e_1\rangle\langle e_0|+\cos2\theta'|e_0\rangle\langle e_1|+(1-\lambda_{000})|e_1\rangle\langle e_1|,\\
      \rho^{E}_{\phi_{000}|c_1^1}=\lambda_{000}|e_0\rangle\langle e_0|-\cos2\theta'|e_1\rangle\langle e_0|-\cos2\theta'|e_0\rangle\langle e_1|+(1-\lambda_{000})|e_1\rangle\langle e_1|,
    \end{split}
  \end{equation}
  respectively. The eigenvalues for $\rho^{E}_{\phi_{000}|c_1^0}$ and $\rho^{E}_{\phi_{000}|c_1^1}$ are the same and expressed as
  \begin{equation}\label{eigenvalues}
    \Lambda_{\pm}=\frac{1}{2}\left(1\pm \sqrt{1-4\lambda_{000}(1-\lambda_{000})+4\lambda_{000}(1-\lambda_{000})\cos2\theta'} \right).
  \end{equation}

  Thus, from Eqs.~\ref{holevo},~\ref{p3E} and ~\ref{eigenvalues}, one can obtain the Holevo quantity
  \begin{equation}
    \begin{split}
      \chi(\rho_{\lambda_{000}})=&h(\{\lambda_{000},1-\lambda_{000}\})-h(\Lambda_{+})\\
       &\leq h(\lambda_{000})
    \end{split}
  \end{equation}
  where the second line is obtained by taking $\Lambda_{+}=1$ with $\theta'=0$, which implies that user $P_3$'s optimal observable to maximize Eve's information is $\sigma_z$ according to Eq.~\ref{basisvec}.

  \item[Step 5] \emph{Analytical Holevo quantity as a function of observed Bell value.}
  From Eq.~\ref{bellentropy}, one can work out that
  \begin{equation}
    \lambda_{000}=\frac{1}{2}\left(1\pm \sqrt{\frac{\langle \hat{B}_{3,\rho_{\lambda_{000}}}\rangle}{4}-4}\right).
  \end{equation}
  Thus, the Holevo quantity $\chi(\rho_{\lambda_{000}})$ given quantum state $\rho_{\lambda_{000}}$ is
  \begin{equation}
    \chi(\rho_{\lambda_{000}})\leq h\left(\frac{1}{2}+\frac{1}{2}\sqrt{\frac{\langle \hat{B}_{3,\rho_{\lambda_{000}}}\rangle}{4}-4} \right).
  \end{equation}
  Similarly, this bound is also applicable to $\rho_{\lambda_{001}}$, $\rho_{\lambda_{010}}$ and $\rho_{\lambda_{011}}$. Thus, we present the bound for the Holevo quantity, which can be saturated by the states in Eq.~\ref{saturate_states},
  \begin{equation}
  \begin{split}
    \chi(A_{31}:E)\leq h\left(\frac{1}{2}+\frac{1}{2}\sqrt{\frac{\langle \hat{B}_{3}\rangle}{4}-4} \right).
  \end{split}
  \end{equation}

  Meanwhile, one can directly generalize the results to more parties considering that there are $N-1$ terms of two-party correlation in Eq.~\ref{optimization}. Thus it holds that
  \begin{equation}
    \chi(A_{N1}:E)\leq h\left(\frac{1}{2}+\frac{1}{2}\sqrt{\frac{\langle\hat{B}_N\rangle}{4}-(N-1)^2}\right).
  \end{equation}

\end{description}

 \begin{table}[h]
  \caption{One-outcome von Neumann entropy $H(A_{31}|E)$ and $H(X|E)$ under given non-maximal entanglement parameter $r$ for this work with $\alpha=1$ and the MABK inequality when $N=3$, respectively.}\label{table:measurement_von}
      \begin{tabular*}{0.415\textwidth}{|c|c|c|c|c|}
        \hline
         $r$ & $\langle \hat{B}_3^{\alpha=1}\rangle$ &$m$ & $H(A_{31}|E)$ (bits) & $H(X|E)$ (bits)\\
         \hline
        $0.05$ & $4.00994$ & $0.564$ & $0.014$ & 0\\
        \hline
        $0.1$ & $4.17612$ & $2.4$ & $0.278$  & 0 \\
        \hline
        $0.14$ & $4.23396$ & $2.7759$ & $0.383$ & 0 \\
        \hline
        $0.145$ & $4.24071$ & $2.81681$ & $0.395$ & $0$ \\
        \hline
        $0.1465$ & $4.24271$ & $2.82885$ & $0.399$ & $2.157*10^{-4}$ \\
        \hline
        $0.15$ & $4.24735$ & $2.85657$ & $0.408$ & $1.45*10^{-2}$ \\
        \hline
        $0.2$ &  $4.30813$ & $3.2$ & $0.531$ & $0.213$ \\
        \hline
        $0.3$ &  $4.4$ & $3.666$ & $0.750$  & $0.571$ \\
        \hline
        $0.4$ &  $4.45421$ & $3.919$ & $0.919$ & $0.856$ \\
        \hline
        $0.5$ &  $4.47214$ & $4$ & $1$  & $1$ \\
        \hline
      \end{tabular*}
      \end{table}

\section{One-outcome von Neumann entropy versus non-maximal entanglement states}\label{vonNeumannVSnonmaximal}

In this part, we present the discussion on the performance for certifying one-outcome von Neumann entropy under imperfections. The first type of imperfect entanglement state is Werner type mixed entanglement state of the form $$\hat{\rho}_w^N=p|GHZ\rangle\langle GHZ|+(1-p)\frac{\hat{I}}{2^N}.$$ Due to the fact that the noise part $\hat{I}$ contributes zero to the Bell operator, the fraction of pure GHZ state should be $p\geq \frac{B_N^{LHV}}{B_3^Q}$ to certify non-zero DI quantum randomness for the Bell inequalities constructed in this work. Thus, it's hard to have better performance compared the Mermin-type inequalities especially when $N$ scales up. For example, for $N=3$ case, the global min entropy can be certified by the Mermin inequality and this work for $p_{Mermin}\geq {1/2}$ and $p\geq \frac{2}{\sqrt{5}}$ with $\alpha=1$, respectively. Meanwhile, for the one-outcome von Neumann entropy, it should be $p_{Mermin} \geq \frac{\sqrt{2}}{2}$ and $p \geq \frac{2}{\sqrt{5}}$ to certify non-zero randomness for the Mermin inequality and this work, respectively, with $\alpha=1$.

Another type of imperfect entanglement state is the pure non-maximal GHZ entanglement state of the form
\begin{equation}\label{non_maximal_state}
  |\Psi\rangle=\sqrt{r}|00\cdots 0\rangle + \sqrt{1-r}|11\cdots 1\rangle
\end{equation}
where $0\leq r \leq 1$ and $r=1/2$ indicates maximal GHZ state. The constructed Bell inequalities can perform better than the MABK inequality. The reason is that non-zero von Neumann entropy can only be certified by the MABK inequality when the Bell values are located beyond the region of genuine multipartite entanglement~\cite{grasselli2021entropy}. Specifically, the corresponding parameter $r$ should be increased to be approximated $r\geq 0.1465$ for the MABK inequality. While, it's $r\geq 0$ for the Bell inequalities constructed in this work to certify non-zero one-outcome von Neumann entropy when taking $\alpha=1$.

The details to obtain above the performance for certifying one-outcome von Neumann entropy versus non-maximal entanglement parameter $r$ with $N=3$ and $\alpha=1$ is an optimization directly over three parties's measurements. The measurement operators for the Bell inequalities constructed by this work is modeled as Eq.~\ref{measurement_this} in the X-Z plane. While, the measurement operators for the MABK inequalities is realized in the X-Y plane as
\begin{equation}\label{measurement_MABK}
\begin{split}
  \hat{A'}_{i0}=\cos\theta_{i0}\sigma_x+\sin\theta_{i0}\sigma_y,\\
  \hat{A'}_{i1}=\cos\theta_{i1}\sigma_x+\sin\theta_{i1}\sigma_y,
\end{split}
\end{equation}
where $\sigma_x$ and $\sigma_y$ are Pauli Matrices, $\theta'_{i0}(\theta'_{i1})$ is an adjustable angle. Thus the Bell values are optimized over the measurements and non-maximal entanglement parameter $r$. Then, recalling the one-outcome von Neumann entropy for this work with $\alpha=1$ and the MABK inequality are $H(A_{31}|E)\geq 1-h\left(\frac{1}{2}+\frac{1}{2}\sqrt{\frac{\langle\hat{B}_3\rangle^2}{4}-4}\right)$ and $H(X|E)\geq 1-h\left(\frac{1}{2}+\frac{1}{2}\sqrt{\frac{m^2}{8}-1}\right)$ respectively, the optimized measurements are as in Table~\ref{table:measurement_von}.

Besides the performance on the one-outcome von Neumann entropy, we also investigated the performance on certifying the global min-entropy versus non-maximal parameter $r$. Specifically, according to the non-maximal entanglement state and measurements in Eq.~\ref{non_maximal_state}, Eq.~\ref{measurement_xz} and Eq.~\ref{measurement_MABK}, we obtain the maximal Bell values for the Bell inequality in this work with $\alpha=10$ and the Mermin inequality by optimization over the measurements for given non-maximal entanglement parameter $r$. Then, with the help of the NPA hierarchy method, one can obtain the guessing probability and global min-entropy bounds for given non-maximal entanglement parameter $r$ and corresponding optimal Bell values. The details can be found in Table~\ref{table:measurement_min_entropy}.

 \begin{table}[h]
  \caption{Global min-entropy $H(A_0B_0C_0|E)$ and $H(A_1B_1C_1|E)$ under given non-maximal entanglement parameter $r$ and corresponding Bell values for this work with $\alpha=10$ and the Mermin inequality when $N=3$, respectively.}\label{table:measurement_min_entropy}
      \begin{tabular*}{0.89\textwidth}{|c|c|c|c|c|c|c|}
        \hline
         $r$ & $\langle \hat{B}_3^{\alpha=10}\rangle$ &$m$& $P_g(A_0B_0C_0|E)$ & $H(A_0B_0C_0|E)$ (bits) & $P_g(A_1B_1C_1|E)$(Ref.~\cite{woodhead2018randomness}) &$H(A_1B_1C_1|E)$ (bits,Ref.~\cite{woodhead2018randomness})\\
         \hline
        $0.05$ & $22.0517$ & $1.7436$ & $0.8317$ & 0.2659&1&0\\
        \hline
        $0.066$ & $22.0669$ & $1.9863$ & $0.7822$  & 0.3543 & 1&0\\
        \hline
        $0.067$ & $22.0678$ & $2.0002$ & $0.7792$ & 0.3599& 0.99996 &$6.3313*10^{-5}$\\
        \hline
        $0.07$ & $22.0706$ & $2.0412$ & $0.7702$ & $0.3767$&0.9893&0.0155 \\
        \hline
        $0.1$ & $22.0971$ & $2.4$ & $0.6840$ & $0.5479$&0.8860&0.1746 \\
        \hline
        $0.2$ & $22.1701$ & $3.2$ & $0.4463$ & $1.1639$& 0.6095&0.7143\\
        \hline
        $0.3$ &  $22.2210$ & $3.6661$ & $0.2805$ & $1.8339$ &0.4027&1.3122\\
        \hline
        $0.4$ &  $22.2511$ & $3.9192$ & $0.1825$  & $2.4541$ &0.2439&2.0357\\
        \hline
        $0.5$ &  $22.2611$ & $4$ & $0.1318$ & $2.9239$& 0.125&3\\
        \hline
      \end{tabular*}
      \end{table}

\end{appendix}






\bibliography{DIRandomness.bib}


%
%
%
%

\end{document}